\documentclass[aps,prd,superscriptaddress,nofootinbib,11pt]{revtex4}
\usepackage[english]{babel}
\usepackage[utf8]{inputenc}
\usepackage{graphicx}   
\usepackage{slashed}
\usepackage{epstopdf}
\usepackage{verbatim}   
\usepackage{color}      
\usepackage{subfigure}  
\usepackage{multirow}
\usepackage{hyperref}   
\usepackage{float}
\usepackage{epsfig,rotating}
\usepackage{amsmath,amssymb}
\usepackage{dsfont}
\restylefloat{table}
\numberwithin{equation}{section}

\newcommand{\vx}{\vec{x}}
\newcommand{\vp}{\vec{p}}
\newcommand{\vq}{\vec{q}}
\newcommand{\vk}{\vec{k}}
\newcommand{\vK}{\vec{K}}

\newcommand{\be}{\begin{equation}}
\newcommand{\ee}{\end{equation}}
\newcommand{\bea}{\begin{eqnarray}}
\newcommand{\eea}{\end{eqnarray}}

\newcommand{\ket}[1]{|#1\rangle}
\newcommand{\bra}[1]{\langle#1|}

\begin{document}
\title{Interaction rates   in  cosmology: \\ heavy particle production and scattering.}

\author{Mudit Rai}
\email{MUR4@pitt.edu} \affiliation{Department of Physics and
Astronomy, University of Pittsburgh, Pittsburgh, PA 15260}
\author{Daniel Boyanovsky}
\email{boyan@pitt.edu} \affiliation{Department of Physics and
Astronomy, University of Pittsburgh, Pittsburgh, PA 15260}

 \date{\today}

\begin{abstract}
We study transition rates and cross sections from first principles in a spatially flat radiation dominated cosmology. We consider a model of scalar particles    to study scattering and heavy particle production   from pair annihilation,   drawing more general conclusions. The   S-matrix formulation is ill suited to study these ubiquitous processes in a rapidly expanding cosmology. We introduce a physically motivated adiabatic expansion that relies  on wavelengths  much smaller than the particle horizon at a given time. The leading order in this expansion dominates the transition rates and cross sections. Several important  and general results are direct consequences of the cosmological redshift and a finite particle horizon: i) a violation of local Lorentz invariance, ii) freeze-out of the production cross section at a finite time, iii) sub-threshold production of heavier particles as a consequence of the uncertainty in the local energy from a  finite particle horizon,   a manifestation of the \emph{antizeno} effect. If heavy dark matter is produced via annihilation of a lighter species, sub-threshold production yields an enhanced abundance. We discuss several possible cosmological consequences of these effects.
\end{abstract}

\keywords{}

\maketitle

\section{Introduction}
\vspace{1mm}
Processes such as scattering and decay play a fundamental role in early Universe cosmology, from post-inflationary reheating\cite{kolb,linde}  to Big Bang Nucleosynthesis (BBN)\cite{kolb,bernstein,dodelson,bbn,fieldsbbn,steigman,sarkar,pos}, and possibly in a successful description of the origin of matter-antimatter asymmetry\cite{trodden,turner,buch,pluma} .  The main approach to studying such phenomena relies on the S-matrix formulation of quantum field theory in Minkowski space time. Within this formulation  a scattering cross section is obtained from the transition probability per unit time from an initial multiparticle state prepared in the infinite past, to another multiparticle state detected in the infinite future normalized  to a unit incoming flux. The cross sections and transition rates obtained in this framework are some of the main ingredients in the kinetic Boltzmann equation that describe the production, evolution and freeze-out of different particle species\cite{kolb,bernstein,dodelson}.
In the S-matrix formulation of transition rates, taking the infinite time limit yields exact energy conservation, and consequently   reaction thresholds.
This approach, when applied to early universe cosmology is at best an approximation, cosmological expansion introduces a distinct time evolution and the different stages, inflation, radiation, matter domination are characterized by different expansion time scales and dynamics. Obviously taking the infinite time limit glosses over the different stages of cosmological expansion  and is in general   unwarranted.
We note that recently, reference\cite{collins} has revisited the S-matrix formulation in Minkowski space time analyzing processes in a finite time interval and discussing in detail the subtleties  of approaching the asymptotic infinite time limit.
Quantum field theory in curved space-time reveals a wealth of unexpected novel phenomena, such as particle production from cosmological expansion \cite{parker,ford,zelstaro,birford,bunch,birrell,fullbook,mukhabook,parkerbook,parfull} along with   processes that are forbidden in Minkowski space time as a consequence of energy/momentum conservation. Pioneering investigations of interacting quantum fields in expanding cosmologies generalized the S-matrix formulation for in-out states in Minkowski space-times for model expansion histories. Self-interacting quantized fields were studied with a focus on renormalization aspects and contributions from pair production to the energy momentum tensor \cite{birford,bunch}.  The decay of a massive particle into two massless particles conformally coupled to gravity was studied in Ref.~\cite{spangdecay} within a modified formulation of the S-matrix for simple cosmological space times.

 Particle decay in   de Sitter space-time  was studied in Refs.~\cite{boydecay,mosca}, revealing surprising phenomena, such as a quantum of a massive field decaying into two (or more) quanta of the \emph{same} field. These phenomena are a direct consequence of the lack of a global  time-like Killing vector, and the concomitant absence of energy conservation. Single  particle decay in an post inflationary cosmology has been studied in refs.\cite{vilja,herringdecay,herringfer} and more recently reheating during a post-inflation period has been studied implementing a Boltzmann equation that inputs cosmological decay rates\cite{vilja2}.
 The results on particle decay of refs.\cite{herringdecay,herringfer} revealed noteworthy   consequences of the cosmological expansion and lack of energy conservation. An important result  for very weakly coupled
 long lived particles is that the Minkowski space-time decay rate underestimates the lifetime of the particle, with potentially important consequences for weakly coupled dark matter.

\vspace{1mm}

\textbf{Motivation and objectives:}
Particle interactions are ubiquitous in cosmology during all the epochs with   fundamental and phenomenological  implications in the description of particle physics processes during the expansion history. While there have been previous studies of single particle decay in inflationary and post inflationary cosmology\cite{spangdecay,vilja,herringdecay,herringfer}, to the best of our knowledge, there has not yet been a systematic study of \emph{cross sections} and interaction rates including consistently the   dynamics of cosmological expansion. Hence, motivated by their  importance   to describe particle physics processes in the early universe, our objectives in this article are the following: {\bf{i:)}} to address the fundamental question on the validity of the S-matrix formulation as applied to cosmology, {\bf{ii:)}} to provide an \emph{ab initio} study of interaction rates and cross sections in a spatially flat radiation dominated (RD) cosmology from a full quantum field theoretical analysis in curved space time, providing a consistent formulation that explicitly includes the cosmological expansion. {\bf{iii:)}} To identify under which circumstances the S-matrix formulation is (approximately) reliable, and when it is not, to establish its limitations. {\bf{iv:)}} To identify potentially new phenomena that is not captured by the S-matrix formulation and that may lead to novel phenomenological consequences.

In this article we begin this program by studying the case of scalar fields with a contact interaction as a first step towards a deeper understanding of fermionic and gauge degrees of freedom of the standard model or beyond and perhaps of possible dark matter candidates. The lessons learned in this study will provide a stepping stone to approaching more general interactions: if they confirm  the validity of an S-matrix approximation to cross sections in cosmology, then our study provides a first principles analysis that lends credibility to this framework, and establishes its limitations. If, on the other hand, our study reveals new phenomena that is not captured by the S-matrix formulation, it may lead to novel phenomenological consequences and will further motivate the study of other interactions relevant to cosmological processes.

\vspace{1mm}

 \textbf{Brief summary of  results:}  After discussing field quantization in a (RD) cosmology, and recognizing the daunting conceptual and technical challenges of obtaining transition matrix elements with the exact field modes even at tree level, we introduce an adiabatic expansion that relies on the ratio $H/E \ll 1$ where $H$ is the expansion rate and $E$ the energy of a particle  measured by a locally inertial observer. This expansion is valid for all wavelengths well inside the particle horizon at any given time and its reliability improves upon cosmological expansion. At heart it hinges on a wide separation between cosmological and microscopic time scales. In this article we consider two   bosonic degrees of freedom with a local contact interaction to leading order in this adiabatic expansion.  We obtain the interaction rate and cross section during the finite time interval determined by the particle horizon.  We   show that the leading adiabatic order yields  the dominant contribution.   As a consequence of the cosmological expansion and a finite particle horizon  we find several  general noteworthy features of the   cross section:
 \textbf{i):} violation of local Lorentz invariance: the cross section for a finite particle horizon features a dependence on the local energy and momenta that explicitly breaks the invariance under local Lorentz transformations.
 \textbf{ii):}    the cross section for heavy particle production features two different phenomena a)   a freeze-out,  whereby the physical momentum falls below the production threshold and the expansion shuts-off the production, b) a time regime during which there is  subthreshold production of a heavier species: the finite particle horizon introduces an energy uncertainty which leads to the  relaxation of the threshold condition and opens a window of width $\propto H$ for production of heavier particles for local energy and momenta that are smaller than the threshold value in Minkowski space-time. This is a manifestation of the \emph{antizeno} effect \cite{antizenokuri,boyaantizeno,antizenoobs}. In other words  the energy uncertainty as a consequence of the finite particle horizon allows processes that would be otherwise   forbidden by strict energy conservation.  A possible implication of this phenomenon may be relevant to dark matter: if heavy dark matter particles are produced via pair annihilation of a much lighter species, sub-threshold production leads to an enhancement of their abundance.

The article is organized as follows: in section (\ref{sec:model}) we introduce the model, discuss the quantization in a spatially flat (RD) cosmology, introduce the adiabatic expansion and obtain the main expressions for the comoving interaction rate and cross section. Section (\ref{sec:massless}) studies the case of massless particles in the final state. This case provides insight into the more general case of massive daughter particles studied in section (\ref{sec:massive}) wherein  violation of local Lorentz invariance  and  the phenomena of freeze-out and subthreshold production as a consequence of the cosmological redshift and finite particle horizon are discussed in detail. In section (\ref{sec:discussion}) we discuss the general lessons and    several important aspects such as a wave packet description, impact on quantum kinetics and BBN.   Section (\ref{sec:conclusions}) summarizes our conclusions and poses further questions.  In appendix (\ref{app:Xminkfiti}) we obtain the cross section for similar processes in Minkowski space-time but during a finite time interval to compare to the results in cosmology.

\section{The model:}\label{sec:model}

In the standard cosmological model, most particle physics processes occur during the radiation dominated (RD) era, therefore we focus  on the post-inflationary   (RD) universe, described by a   spatially flat Friedmann-Robertson-Walker (FRW) cosmology  with the metric in comoving coordinates given by
\be  g_{\mu \nu} = \textrm{diag}(1, -a^2, -a^2, -a^2) \,, \label{frwmetric}\ee where $a$ is the scale factor. It is convenient to pass    to conformal time $\eta$ with $d\eta = dt/a(t)$, in terms of which the metric becomes ($a(t)\equiv a(\eta)$)
\be  g_{\mu \nu} = a^2(\eta)\,\eta_{\mu \nu} \, . \label{conformalmetric} \ee where $\eta_{\mu \nu}$ is the Minkowski space-time metric.

 During the     (RD)  stage, the Hubble expansion rate $H(\eta)$ is given by
\be H(\eta) = \frac{1}{a^2(\eta)}\frac{d a(\eta)}{d\eta} = 1.66 \sqrt{g_{eff}}\,\frac{T^2_0}{M_{Pl}\,a^2(\eta)}\,, \label{hrd}\ee where $g_{eff}$ is the effective number of ultrarelativistic degrees of freedom, which varies in time as different particles become non-relativistic, and $T_0$ is the temperature of the cosmic microwave background today. We take $g_{eff}=2$ corresponding to radiation today, which yields a lower bound on the Hubble scale,  in order to estimate   order of magnitudes.  The scale factor during (RD) is
\be a(\eta)= H_R   ~\eta ~~;~~ H_R= H_0\,\sqrt{\Omega_R} = 10^{-44}\,\mathrm{GeV} \Rightarrow H(\eta)= \frac{H_R}{a^2(\eta)}\,,  \label{aofetarm}\ee ($H(\eta)\equiv H(t)$) where quantities with a subscript $0$ refer to the values today, and $\Omega_R$ is the ratio of energy density in radiation today to the critical density. The numerical value for $H_R$ follows from taking $g_{eff}=2$. During this stage  the relation between conformal and comoving time is given by \be \eta = \Big( \frac{2\,t}{H_R}\Big)^{\frac{1}{2}} \Rightarrow a(t) = \Big[ 2\,t H_R\Big]^{\frac{1}{2}}~~;~~ H(t)= \frac{1}{2\,t}\,. \label{etaoft} \ee

If particles interact directly with the thermal bath that constitutes the environment during the (RD) era, high temperature and density induce thermal corrections to masses and interaction vertices. In this article we do not consider these effects, focusing solely on the conceptual aspects of obtaining interaction rates and cross sections directly in real time accounting for the cosmological expansion.  Therefore, the  results obtained are  general and independent of the thermal aspects of populations, and apply directly, for example to two dark matter particle species that interact weakly with each other via a local contact interaction but not with the standard model degrees of freedom. Including finite temperature and density corrections to these quantities in the case of direct interaction and thermalization  with the environmental plasma remains  a longer term goal outside the scope of this article.

We consider a complex   ($\Psi$) and a real ($\Phi$) scalar field  with a local quartic interaction  with action given by
\be  A   =    \int d^4 x \sqrt{|g|} \Bigg\{ g^{\mu\nu}\,\partial_\mu  \Psi^\dagger \partial_\nu \Psi-M^2 \Psi^\dagger \Psi+ \frac{1}{2} g^{\mu\nu}\,\partial_\mu \Phi \partial_\nu \Phi-\frac{1}{2}  m^2 \Phi^2 -\lambda \, \Psi^\dagger\,\Psi\,\Phi^2\,  \Bigg\}
\label{action}\ee noting  that the Ricci scalar vanishes identically in a spatially flat (RD)- (FRW) cosmology. In comoving spatial coordinates and   conformal time and upon conformally rescaling the fields as
\be \Psi(\vec{x},t) = \frac{\phi(\vec{x},\eta)}{a(\eta)} ~~;~~\Phi(\vec{x},t) = \frac{\chi(\vec{x},\eta)}{a(\eta)}~~;~~ a(\eta) = a(t(\eta))\,, \label{rescale}\ee   the action (\ref{action}) becomes
\begin{align}
A=& \int d^3x \, d\eta  \biggl\{\tfrac{1}{2}  \,\Big(\frac{d\chi}{d\eta} \Big)^2 -\tfrac{1}{2} \,\bigl( \nabla \chi\bigr)^2 -\tfrac{1}{2} \chi^2\,  m^2\,a^2(\eta) \\
& + \Big(\frac{d\phi^\dagger}{d\eta} \Big)\,\Big(\frac{d\phi}{d\eta} \Big) -\bigl( \nabla \phi^\dagger\bigr) \,\bigl( \nabla \phi\bigr) - |\phi|^2\, M^2\,a^2(\eta) -\lambda \, \phi^\dagger(\vx,\eta)\,\phi(\vx,\eta)\,\chi^2(\vx,\eta) \,  \biggr\}\,.
\end{align}

Although this simple model cannot capture all of the important aspects of standard model physics and/or dark matter candidates,   it allows us to study ubiquitous phenomena such as scattering  and particle production, illuminating the interplay between the dynamics of expansion and threshold kinematics. Furthermore, its analysis  leads us to  draw general lessons on the technical and conceptual aspects that will pave the way towards understanding   interactions more relevant to particle physics models.

\vspace{1mm}

\subsection{Quantization and adiabatic expansion:}\label{sub:quantad} We begin with the quantization of free fields \cite{parker,ford,zelstaro,birrell,fullbook,parkerbook,mukhabook,birford,bunch}   as a prelude to the interacting theory. The Heisenberg equations of motion for the conformally rescaled fields $\phi, \chi$ in conformal time are
\bea \frac{d^2}{d\eta^2}\,\phi(\vec{x},\eta) - \nabla^2 \phi(\vec{x},\eta) + M^2\,a^2(\eta)\,\phi(\vec{x},\eta) &  = & 0 \,,\label{Eomphi}\\\frac{d^2}{d\eta^2}\,\chi(\vec{x},\eta) - \nabla^2 \chi(\vec{x},\eta) + m^2\,a^2(\eta)\,\chi(\vec{x},\eta) &  = & 0 \,. \label{Eomchi}
\eea
It is convenient to quantize the fields  in a comoving volume $V$, namely,
\be   \phi(\vx,\eta) =  \frac{1}{\sqrt{ V}}\, \sum_{\vk}\Big[ a_{\vk} \,g_k(\eta)\, e^{i\vk\cdot\vx} + b^\dagger_{\vk} \,g^*_k(\eta) \, e^{-i\vk\cdot\vx}\Big]\,. \label{fige} \ee
\be   \chi(\vx,\eta) =  \frac{1}{\sqrt{ V}}\, \sum_{\vk}\Big[ c_{\vk} \,f_k(\eta)\, e^{i\vk\cdot\vx} + c^\dagger_{\vk} \,f^*_k(\eta) \, e^{-i\vk\cdot\vx}\Big]\,,  \label{chief} \ee
 where the mode functions $g_k(\eta);f_k(\eta)$ are solutions of the following equations
 \bea \Big[ \frac{d^2}{d\eta^2}+ \Omega^2_k(\eta)   \Big]g_k(\eta)  & = & 0 ~~;~~ \Omega^2_k(\eta)= k^2 + M^2\,a^2(\eta) \label{equfi} \\
 \Big[ \frac{d^2}{d\eta^2}+ \omega^2_k(\eta)  \Big]f_k(\eta)  & = & 0 ~~;~~ \omega^2_k(\eta) = k^2+ m^2 a^2(\eta) \,, \label{equchi}\eea  and satisfy   the Wronskian condition
\begin{align}
& g^{\,'}_k(\eta)g^*_k(\eta) - g^{*\,'}_k(\eta) g_k(\eta) = -i \,  \\
& f^{\,'}_k(\eta)f^*_k(\eta) - f^{*\,'}_k(\eta) f_k(\eta) = -i \,, \label{wronski}
\end{align}
so that the annihilation   and creation   operators are \emph{time independent} and obey the canonical commutation relations $[a_{\vk},a^{\dagger}_{\vk'}] = \delta_{\vk,\vk'} \,; [c_{\vk},c^{\dagger}_{\vk'}] = \delta_{\vk,\vk'}\,\mathrm{etc.} $, and the vacuum state $\ket{0}$ is defined as
\be a_{\vk}\ket{0}= 0~;~b_{\vk}\ket{0}= 0 ~;~ c_{\vk}\ket{0}= 0 \,. \label{vaccu}\ee

Since the mode equations for $f_k(\eta),g_k(\eta)$ obey  similar equations of motion, we focus on $f_k(\eta)$ from which we can obtain $g_k(\eta)$ by the replacement $m \rightarrow M$.
Let us introduce the dimensionless variables
\be x = \sqrt{2m\,H_R} ~ \eta ~~;~~ \alpha = -\frac{k^2}{2m\,H_R}\,, \label{weberparas} \ee in terms of which the equation (\ref{equchi}) is identified with Weber's equation\cite{gr,as,nist,bateman,magnus}
\be \frac{d^2}{dx^2}\,f(x) +\Big[\frac{x^2}{4}-\alpha \Big]f(x) =0 \,.\label{webereq}\ee The general solutions are  linear combinations of Weber's parabolic cylinder functions $W[\alpha;\pm x]$\cite{gr,as,nist,bateman,magnus}. We seek solutions that can be identified with   particle states obeying the condition \be f_k(\eta) \rightarrow  \frac{e^{-i\,\int^{\eta}_{\eta^*}\,\omega_k(\eta')\,d\eta'}}{\sqrt{2\,\omega_k(\eta)}}\,, \label{outstates}\ee for wavevectors well inside the particle horizon (Hubble radius) which is discussed below in more detail,  along with the Wronskian condition (\ref{wronski}). The lower limit $\eta^*$ corresponds to a conformal time at which the condition of adiabaticity $\omega'_k(\eta)/\omega^2_k(\eta) \ll 1$, described in detail below (see eqns. (\ref{firstordad},\ref{adpara}))  is fulfilled. These were found in ref.\cite{herringdm}, and are given by
\be f_k(\eta) = \frac{1}{(8mH_R)^{1/4}}\,\Big[\frac{1}{\sqrt{\kappa}}\,W[\alpha;x] -i\sqrt{\mathcal{D}}\,W[\alpha;-x]  \Big]~~;~~ \mathcal{D} = \sqrt{1+e^{-2\pi|\alpha|}}-e^{-\pi|\alpha|}\,.\label{fconf}\ee It is shown in ref.\cite{herringdm} that the asymptotic behavior of $f_k(\eta)$   is indeed given by (\ref{outstates}) for wavelengths much smaller than the particle horizon as well as  in the long time limit. The mode functions $g_k(\eta)$ are obtained from these by the replacement $m \rightarrow M$.

 A perturbative approach to obtaining a cross section begins by defining the interaction picture wherein fields feature the free field time evolution (\ref{fige},\ref{chief}) with the mode functions obeying the free field equations of motion (\ref{equfi},\ref{equchi}). The transition matrix elements between initial and final states are obtained in a perturbative expansion in terms of time ordered integrals of products of the interaction Hamiltonian in the interaction picture.  Even to lowest order in such perturbative expansion, the transition matrix elements would feature time and momentum integrals of products of the mode functions $f_k,g_k$ given by eqn. (\ref{fconf}) and similarly for $g_k$. Obviously this is very different from the situation in Minkowski space-time where the mode functions are simple plane wave solutions both in space and time, and  integrals over the space-time coordinates lead to energy momentum conservation at each vertex. In a spatially flat (FRW) cosmology   there are three space-like Killing vectors associated with spatial translational invariance, therefore  the spatial part of the mode functions are the usual plane waves, as manifest in the expansions (\ref{fige},\ref{chief}). Hence,  the spatial integrals yield spatial momentum conservation, however, the time integrals involve   products
 of parabolic Weber functions, obviously presenting a formidable technical obstacle. Above and beyond these technical difficulties, it is   clear that unlike Minkowski space-time, there is no energy conservation, this is consequence of the lack of a time-like Killing vector in an expanding cosmology. Most particle physics processes in the early Universe are deemed to involve energetic particles, which motivates us to invoke an adiabatic approximation for the mode functions based on the well studied Wentzel-Kramers-Brillouin (WKB) approximation to the solution of the equations for the mode functions\cite{birrell,fullbook,mukhabook,parkerbook,birford,bunch}.

Since both mode functions satisfy the same differential equations, albeit with a different mass term, we will carry out the WKB analysis for $f_k(\eta)$. Writing the solution of the mode equations in the WKB form\cite{birrell,fullbook,mukhabook,parkerbook,birford,bunch,dunne,wini}
\be f_k (\eta) = \frac{e^{-i\,\int^{\eta}_{\eta_i}\,W_k(\eta')\,d\eta'}}{\sqrt{2\,W_k(\eta)}} \,, \label{WKB}\ee and inserting this ansatz into (\ref{equchi}) it follows that $W_k(\eta)$ must be a solution of the equation\cite{birrell}
\be W^2_k(\eta)= \omega^2_k(\eta)- \frac{1}{2}\bigg[\frac{W^{''}_k(\eta)}{W_k(\eta)} - \frac{3}{2}\,\bigg(\frac{W^{'}_k(\eta)}{W_k(\eta)}\bigg)^2 \bigg]\,.  \label{WKBsol} \ee

This equation can be solved in an \emph{adiabatic expansion}
\be W^2_k(\eta)= \omega^2_k(\eta) \,\bigg[1 - \frac{1}{2}\,\frac{\omega^{''}_k(\eta)}{\omega^3_k(\eta)}+
\frac{3}{4}\,\bigg( \frac{\omega^{'}_k(\eta)}{\omega^2_k(\eta)}\bigg)^2 +\cdots  \bigg] ~~;~~ \omega_k(\eta) = \sqrt{k^2+m^2a^2(\eta)} \,.\label{adexp}\ee We refer to terms that feature $n$-derivatives of $\omega_k(\eta)$ as of $n$-th adiabatic order. The nature and reliability of the adiabatic expansion is revealed by considering the term of first adiabatic order for generic mass $m$:
\be \frac{\omega^{'}_k(\eta)}{\omega^2_k(\eta)} = \frac{m^2\, a(\eta) a^{'}(\eta)}{\Big[ k^2 + m^2\,a^2(\eta) \Big]^{3/2}}\,, \label{firstordad}\ee this is most easily recognized in \emph{comoving} time $t$, introducing the \emph{local} energy $E_k(t)$ and Lorentz factor $\gamma_k(t)$ measured by a comoving observer in terms of the \emph{physical} momentum $k_p(t) = k/a(t)$
\bea E_k(t) & = &  \sqrt{k^2_p(t)+m^2} = \frac{\omega_k(\eta)}{a(\eta)}  \label{comoener}\\
 \gamma_k(t) & = & \frac{E_k(t)}{m} \,,\label{gamafac} \eea  and the Hubble expansion rate
 $H(t) = \frac{\dot{a}(t)}{a(t)} = a^{'}/a^2 $.    In terms of these variables, the first order adiabatic ratio  (\ref{firstordad}) becomes
 \be \frac{\omega^{'}_k(\eta)}{\omega^2_k(\eta)} = \frac{H(t)}{\gamma^2_k(t)\,E_k(t)}\,.  \label{adratio}\ee

 In similar fashion the higher order terms in the adiabatic expansion for a (RD) cosmology (vanishing Ricci scalar) can be obtained,
\begin{align}
\frac{\omega^{''}_k(\eta)}{\omega_k^3(\eta)} &= \frac{1}{\gamma^2_k(t)} \frac{H^2(t)}{E_k^2(t)}\Big[1-\frac{1}{\gamma^2_k(t)}\Big] \nonumber\\
\frac{\omega^{'''}_k(\eta)}{\omega_k^4(\eta)} &= -  \frac{3}{\gamma^3_k(t)}\,\frac{H^3}{E^3_k} \Big[1-\frac{1}{\gamma^2_k(t)} \Big] \,. \label{secthirdad}
\end{align}
   Consequently, (\ref{adexp}) takes the form:
\begin{equation}
W^2_k(t) = a^2(t)E^2_k(t)\Big[1-\frac{1}{2\gamma^2_k(t)} \frac{H^2(t)}{E_k^2(t)}\Big[ 1-\frac{5}{2\gamma^2_k(t)}\Big] +\cdots \Big]\,.
\end{equation}

Since the Ricci scalar ($R \propto a''(\eta)/a^3(\eta)$) vanishes in a (RD) cosmology, it follows that for $m=0$ ($\gamma_k = \infty$) the mode functions are the same as in Minkowski space-time and the WKB approximation becomes exact, furthermore, for $k=0$ ($\gamma_k =1$) only $\omega'_k/\omega^2_k$ remains since higher order derivatives of the frequency ($\omega_{k=0}(\eta)=m\,a(\eta)$) vanish.

These two limits and the expansion terms featured above lead us to   identify  the   ratio
\be  \frac{H(t)}{\gamma_k(t)\,E_k(t)} \ll 1 \,, \label{adpara}\ee as the small, dimensionless \emph{adiabatic} expansion parameter.  We will instead adopt a more stringent condition for the adiabatic approximation, namely
\be \frac{H(t)}{E_k(t)} \ll 1 \Rightarrow E_k(t)\,t \gg 1\,,  \label{adiarat} \ee where we used the relation (\ref{etaoft}) in the second inequality.

  The physical interpretation of the   ratio $H(t)/E_k(t)$ is clear: typical particle physics degrees of freedom feature either physical de Broglie or Compton wavelengths that are much smaller than the (physical) particle horizon $\propto 1/H(t)$   at any given time during (RD). In a standard (RD) cosmology the particle horizon always grows faster than a physical wavelength, therefore the reliability of the adiabatic expansion improves with the cosmological expansion. The condition (\ref{adiarat}) is also equivalent to a ``long time limit'' in the sense that there are many oscillations of the microscopic degrees of freedom during a Hubble time scale $\simeq 1/H(t)$. 

As an example, let us consider processes occurring early  in the (RD) stage, for example   at the Grand Unification (GUT)  scale  $\simeq 10^{15}\,\mathrm{GeV}$, assuming that particles feature  \emph{physical} momenta at this scale $k_{ph}(\eta) = k/a(\eta) \simeq 10^{15}\,\mathrm{GeV}$ with $k$ being the \emph{comoving} momentum and a mass $\simeq 100\,\mathrm{GeV}$, hence a local Lorentz factor $\gamma_k \simeq 10^{13}$. If the environmental temperature of the plasma is $T \simeq T_{\text{GUT}} \simeq 10^{15}\,\mathrm{GeV}$ and taking as an example the standard model result $g_{eff} \simeq 100$, it follows that $H \simeq 10^{12}\,\mathrm{GeV}$ and approximating
$T_{\text{GUT}} \simeq T_\text{CMB}/a(\eta_i)$ implying that the scale factor at the GUT scale $a(\eta_i) \simeq 10^{-28}$  and a  \emph{comoving} wavevector    $k \simeq 10^{-13}\,\mathrm{GeV}$. This situation yields a  ratio $H/E \simeq 10^{-3}$, which becomes smaller with the cosmological expansion and the adiabatic ratio is even much smaller on  account of the Lorentz factor.


Although the value chosen for the scale factor $a\simeq 10^{-28}$ is probably near the end
of inflation, the main point is that even for these values the adiabatic ratio $H/E$ is small enough that the adiabatic approximation is reliable. As the cosmological expansion proceeds this ratio becomes even smaller, making the adiabatic approximation even more reliable. In conclusion, the assumption of the validity of the adiabatic approximation for wavelengths smaller than the Hubble radius is reliable all throughout the (RD). Hence, our analysis, based on this approximation, is valid during this era of cosmological expansion.


Underlying this analysis of energy scales, the adiabatic approximation entails a wide separation of \emph{time scales}: the expansion time scale $1/H$ is much longer than the microscopic time scales of oscillations associated with particle states $\simeq 1/E$ as implied by the inequality(\ref{adiarat}). It is this separation of time scales that warrants the adiabatic approximation and will undergird our analysis below.

This analysis clarifies that the adiabatic approximation breaks down for  non-relativistic particles $\gamma_k \simeq 1$ with masses $m \leq H$. This situation corresponds to $k_{ph} \ll m \ll H$, hence the breakdown of the adiabaticity condition is associated with wavelengths that are larger than the particle horizon at a given time.  However, since in a (RD) cosmology the particle horizon grows    $\propto a^2(\eta)$ whereas physical wavelengths grow   $\propto a(\eta)$ and the Compton wavelengths remain constant, eventually a superhorizon mode enters the particle horizon and the adiabatic approximation eventually becomes reliable. This analysis delineates the regime  of applicability of the adiabatic approximation and its implementation must always be accompanied with an analysis of the relevant scales.

In this article we focus our study on obtaining the cross section to leading orders in the coupling and the adiabatic expansion, the latter implies  the \emph{zeroth-adiabatic} order for mode functions, namely
\be f_k (\eta) = \frac{e^{-i\,\int^{\eta}_{\eta_i}\,\omega_k(\eta')\,d\eta'}}{\sqrt{2\,\omega_k(\eta)}}~~;~~ g_k(\eta) = \frac{e^{-i\int^\eta_{\eta_i} \Omega_k(\eta')\,d\eta'}}{\sqrt{2\,\Omega_k(\eta)}} \,. \label{WKBzo}\ee

 It is shown explicitly in ref.\cite{herringdm} that the exact mode functions given by eqn. (\ref{fconf}) coincide with zeroth adiabatic order $f_k(\eta)$ given by eqn. (\ref{WKBzo}) to leading order  in the adiabatic expansion $\omega'(\eta)/\omega^2(\eta)\ll 1$ considered in this study.

 We will show explicitly that higher order adiabatic corrections provide small contributions to the  processes considered here which are further suppressed by the coupling. The phase of the mode function has an immediate interpretation in terms of comoving time and the local comoving energy (\ref{comoener}), namely
\be e^{-i\,\int^{\eta}_{\eta_i}\,\omega_k(\eta')\,d\eta'} = e^{-i\,\int^{t}_{t_i}\,E_k(t')\,dt'}\,.\label{phase}\ee
where we used the relations $\omega_k = a(\eta) E_k ~~;~~ a(\eta)d\eta = dt$. A similar analysis holds for the phases involving $\Omega_k(\eta)$. This  is a natural and straightforward generalization of the phase of \emph{positive frequency particle states in Minkowski space-time} which  is precisely the rational for the   boundary conditions (\ref{outstates}) on the mode functions\cite{herringdm}.


 Since as shown in ref.\cite{herringdm} to leading order in the adiabatic expansion the mode functions $f_k(\eta),g_k(\eta)$ in (\ref{fige},\ref{chief}) are given by (\ref{WKBzo}),   the expansion of the $\phi,\chi$ fields (\ref{fige},\ref{chief}) to leading adiabatic order  become
\be \phi(\vx,\eta) = \sum_{\vk} \frac{1}{\sqrt{2\,V\,\Omega_k(\eta)}}\,\Big[ a_{\vk} \,e^{-i\int^\eta_{\eta_i} \,\Omega_k(\eta')\,d\eta'}\, e^{i\vk\cdot\vx} + b^\dagger_{\vk} \,e^{ i\int^\eta_{\eta_i} \,\Omega_k(\eta')\,d\eta'}\, e^{-i\vk\cdot\vx}\Big]\,, \label{fiexp} \ee

\be \chi(\vx,\eta) = \sum_{\vk} \frac{1}{\sqrt{2\,V\,\omega_k(\eta)}}\,\Big[ c_{\vk} \,e^{-i\int^\eta_{\eta_i} \,\omega_k(\eta')\,d\eta'}\, e^{i\vk\cdot\vx} + c^\dagger_{\vk} \,e^{ i\int^\eta_{\eta_i} \,\omega_k(\eta')\,d\eta'}\, e^{-i\vk\cdot\vx}\Big]\,, \label{chiexp} \ee with the vacuum state $\ket{0}$ obeying the condition (\ref{vaccu}).

We will refer to the Fock states created out of the vacuum  for the fields $\phi^\dagger~;~\phi$ as $\varphi^{\pm}$ particle-antiparticle respectively and  $\chi$ particles for the $\chi$ fields, where the particle interpretation is warranted by the form of the zeroth-order adiabatic solutions (\ref{WKBzo}).
The Minkowski space-time limit is obtained by setting $a(\eta)=1$ and $\eta \rightarrow t$, so that the frequencies become time independent, and  absorbing the dependence on the initial time $t_i$ into a (phase) redefinition of the creation and annihilation operators.

In order to establish a clear identification of the zeroth order adiabatic modes with particles we analyze the free-field Hamiltonian, which in terms of the conformally rescaled field operators and focusing on the $\chi$ field is given by
\begin{equation}
 H_{\chi}(\eta) =\frac12\int d^3x\,\{ \pi^2+(\nabla\chi)^2+m^2 a^2(\eta)\chi^2\}~~;~~ \pi \equiv \chi' \,.
\end{equation}
Using the canonical commutation relations, and the Wronskian conditions (\ref{wronski}) it is straightforward to find that the Heisenberg equations of motion obtained with this Hamiltonian are precisely eqn. (\ref{equfi}) for the mode functions $f_k$, hence for the quantum field $\chi$. Using the expansion (\ref{chief}) for the $\chi$ field and integrating in $d^3x$ we find
\begin{align}
  H_{\chi}(\eta)  = \frac{1}{2}\sum_{\vec{k}}\Bigl{\{}
    c^\dagger_{\vec{k}}  c_{\vec{k}} \, \Big[|f'_k|^2+\omega^2_k(\eta)\,|f_k|^2\Big]
    +\Big(c_{\vec{k}} c_{-\vec{k}}\,\Big[(f'_k)^2+\omega^2_k(\eta)(f_k)^2\Big] + h.c.\Big)
    \Bigr{\}}\,. \label{Hchi}
\end{align} Writing $f_k(\eta)$ in the WKB form (\ref{WKB}) and keeping the zeroth adiabatic order, we find
\be H_{\chi}(\eta)= \sum_{\vec{k}}
    c^\dagger_{\vec{k}}  c_{\vec{k}} \,\omega_k(\eta)\,. \label{H0chi} \ee
A similar analysis for the $\phi$ field to zeroth order in the adiabatic expansion yields
\be H_{\phi}(\eta)= \sum_k
    \Big(a^\dagger_{\vec{k}}  a_{\vec{k}}  + b^\dagger_{\vec{k}} b_{\vec{k}}\Big)\Omega_k(\eta)\,. \label{H0fi} \ee It is straightforward to confirm that the terms   $c_{\vec{k}} c_{-\vec{k}}$ in (\ref{Hchi}) (and similar for the $\phi$ field)   are of second and higher adiabatic order\cite{herringdecay}.

    \subsection{Amplitudes and rates:}\label{subsec:rates}

    Since we will study transition rates in a finite time interval it is important to address their proper definition. In S-matrix theory a rate is simply defined by taking the transition probability from $t=-\infty$ to $t= +\infty$ and dividing by the total time interval. In this infinite time limit the total transition probability grows linearly in time, therefore dividing by the total time elapsed yields a time independent transition \emph{rate}. However, during a finite time interval the transition probability features a more subtle time dependence and a consistent definition of the transition rate must be carefully reassessed.  Reference\cite{herringdecay} introduced a formulation of the time evolution of states in  cosmology which   leads to an   identification of transition probabilities and rates. Such a formulation was also implemented in ref.\cite{boyaantizeno} to study the dynamics of decay directly in real time in Minkowski space-time. For consistency of presentation we summarize the main aspects of this formulation here in order to clarify the definition of transition rates which are the main focus of this study. We refer the reader to these references for more details.
    In the interaction picture states evolve as
    \be |\Psi(\eta)\rangle_I = U_I(\eta,\eta_0) \,|\Psi(\eta_0)\rangle_I \,, \label{psiint}\ee where the time evolution operator in the interaction picture obeys
    \begin{equation}
i\frac{d}{d\eta}U_I(\eta,\eta_0) = H_I(\eta)U_I(\eta,\eta_0)   ~~;~~ U_I(\eta_0,\eta_0) =1\,.  \label{Uip}
\end{equation} whose perturbative solution yields
    \be U_I(\eta;\eta_i) = 1 - i \int^\eta_{\eta_i} H_I(\eta')\,d\eta' +\cdots \label{Uop}\ee
where
\be H_I(\eta) = \lambda \int d^3x \phi^\dagger(\vx,\eta) \phi(\vx,\eta) \chi^2(\vx,\eta)\,,\label{Hint}\ee is the interaction Hamiltonian in the interaction picture. Expanding $|\Psi(\eta)\rangle_I$ in the adiabatic Fock states eigenstates of $H_{\chi},H_{\phi}$ generically labeled as $|n\rangle$,  as $\ket{\Psi(\eta)}_I = \sum_n C_n(\eta)\ket{n}$ the expansion coefficients (amplitudes) obey
\begin{equation}
i\frac{d}{d\eta} C_n(\eta) = \sum_m C_m(\eta)\bra{n}H_I(\eta)\ket{m}\,.
\end{equation}

In principle this is an infinite hierarchy of integro-differential equations for the coefficients $C_n(\eta)$; progress can be made, however, by considering states connected
by the interaction Hamiltonian to a given order in the interaction. Consider that initially the state is $\ket{A}$ so that $C_A(\eta_i) =1\,;\,C_{\kappa}(\eta_i) =0$ for $\ket{\kappa}\neq \ket{A}$, and consider a  first order transition  process
$\ket{A}\rightarrow\ket{\kappa}$  to    intermediate multiparticle states $\ket{\kappa}$ with transition matrix elements $\langle \kappa|H_I(\eta)|A\rangle$. Obviously the state $\ket{\kappa}$ will be connected to other multiparticle states $\ket{\kappa'}$  different from $\ket{A}$ via $H_I(\eta)$. Hence for example up to second order in the interaction, the  state $\ket{A}\rightarrow \ket{\kappa}\rightarrow \ket{\kappa'}$. Restricting the hierarchy to \emph{first order transitions} from the initial state $\ket{A} \leftrightarrow \ket{\kappa}$ results in a coupled set of equations
\bea
i\frac{d}{d\eta} C_A(\eta) &= &  \sum_\kappa C_\kappa(\eta)\bra{A}H_I(\eta)\ket{\kappa}~~;~~ C_A(\eta_i) =1\,,\label{eqA}\\
i\frac{d}{d\eta} C_\kappa(\eta) &= & C_A(\eta)\bra{\kappa}H_I(\eta)\ket{A}~~;~~ C_{\kappa}(\eta_i) =0 \,. \label{eqK}
\eea The hermiticity of $H_I$ leads to the result
\be \frac{d}{d\eta}\,\Bigg\{|C_A(\eta)|^2 + \sum_\kappa |C_\kappa(\eta)|^2 \Bigg\} = 0  \Rightarrow |C_A(\eta)|^2 + \sum_\kappa |C_\kappa(\eta)|^2=1\,, \label{totder}\ee this is the statement of unitarity and we have used the initial conditions on the amplitude.
For the cases under consideration, for example that of pair annihilation $\varphi^+ \varphi^- \rightarrow \chi \chi$ studied below, the initial state $|A\rangle$ is the two particle state $|A\rangle = \ket{1^+_{\vk_1};1^{-}_{\vk_2}}$ namely with a particle-antiparticle pair of momenta $\vk_1;\vk_2$ respectively and the states $|\kappa \rangle$ being a final state of two $\chi$ particles, namely  $|\kappa \rangle = \ket{1_{\vp_3};1_{\vp_4}} $.

Following ref.\cite{herringdecay} we solve the system of equations (\ref{eqA},\ref{eqK}) to leading order in the interaction consistently with our tree level calculation of the transition amplitude, yielding
\be
\frac{d}{d\eta} C_A(\eta) = -\int_{\eta_i}^\eta d\eta' \,
\Sigma_A(\eta,\eta')~~ C_A(\eta)~~;~~ C_A(\eta_i)=1\,,\label{marko} \ee
\be C_{\kappa}(\eta) = -i\int^{\eta}_{\eta_i}\,\bra{\kappa}H_I(\eta')\ket{A}\,C_{A}(\eta')\,d\eta' \,, \label{kappapop}\ee where the self-energy
\be \Sigma_A(\eta;\eta') =
\sum_\kappa \bra{A}H_I(\eta)\ket{\kappa}
\bra{\kappa}H_I(\eta')\ket{A} \,. \label{sigmaA}\ee
The solution of eqn. (\ref{marko}) yields the probability of remaining in the initial state $|A\rangle$ as
\be  \mathcal{P}_A(\eta)= |C_A(\eta)|^2 = e^{-\int^\eta_{\eta_i} \Gamma_A(\eta') d\eta'}   ~~;~~ \Gamma_A(\eta) = 2 \int^\eta_{\eta_i}d\eta_1\,\mathrm{Re}\,[\Sigma_A(\eta,\eta_1)]   \,, \label{probbaA} \ee and the transition rate
\be \frac{1}{\mathcal{P}_A(\eta)} \frac{d}{d\eta} \mathcal{P}_A(\eta) = -\Gamma_A(\eta) \,. \label{tranrat}\ee To leading order in the interaction, using the unitarity condition  (\ref{totder})  we find that the total probability of final states is given by
\be \mathcal{P}_{tot}(\eta) = \sum_\kappa |C_\kappa(\eta)|^2 = \int^\eta_{\eta_i} \Gamma_A(\eta')\, d\eta' =  2\,\int^\eta_{\eta_i} d\eta_2 \int^{\eta_2}_{\eta_i} d\eta_1 \mathrm{Re}\Sigma_A[\eta_2;\eta_1]\,,  \label{Ptot} \ee
hence
\be \frac{d}{d\eta}\mathcal{P}_{tot}(\eta) = \Gamma_A(\eta)\,. \label{derptot}\ee
This is a manifestation of  the optical theorem in real time which licenses us to  \emph{define} the  comoving transition \emph{rate} as
\be \Gamma_{i\rightarrow f}(\eta) = \frac{ d }{d\eta} \mathcal{P}_{tot}(\eta) = \Gamma_A(\eta) = 2\, \int^{\eta}_{\eta_i} d\eta_1 \mathrm{Re}\Sigma_A[\eta;\eta_1]\,.\label{rate}\ee

In Minkowski space-time in the long time limit $t\rightarrow \infty$  the rate $\Gamma_A$ becomes independent of time and at long time $\mathcal{P}_A = e^{-\Gamma_A t}$  while the total probability to final states $\mathcal{P}_{tot} = \Gamma_A\,t $.  Therefore, in this case   defining the rate as $\mathcal{P}_{tot}/t$ yields the same result as $d \mathcal{P}_{tot}/dt$. In cosmology where the background is time dependent and momenta are redshifted by the expansion, or during a finite time interval during which the linear growth in time of the final total probability is not yet established,  the two different definitions obviously yield different results. This indicates the subtleties associated with definitions of transition rates in a finite time interval.

The total number of ``events'' is given by $\mathcal{P}_{tot}(\eta)$ namely the (conformal) time integral of the transition rate, which is manifestly positive.
Therefore, motivated by the above analysis based on  unitarity and the optical theorem we \emph{define} the transition \emph{rate} as in eqn. (\ref{rate}). Some subtleties associated with this definition, and an analysis of an alternative definition closer to that in Minkowski space- time and its caveats will be discussed in section (\ref{sec:discussion}).

\subsection{Pair annihilation: $\varphi^+ \varphi^- \rightarrow \chi \chi$:  }
 For this process
   \be \ket{i} = \ket{1^+_{\vk_1};1^{-}_{\vk_2}} ~~;~~ \ket{f} = \ket{1_{\vp_3};1_{\vp_4}} \label{ifstates} \,,\ee these are eigenstates of the zeroth order adiabatic Hamiltonians (\ref{H0chi},\ref{H0fi}),  and the transition matrix element $\langle f|H_I(\eta)|i\rangle$ is given by
\be   \langle f|H_I(\eta)|i\rangle  =     \frac{-i\,\lambda}{2\,V}\,\delta_{\vk_1+\vk_2 - \vp_3 -\vp_4}~~ \frac{e^{-i \int^{\eta}_{\eta_i} \big(\Omega_{k_1}(\eta')+\Omega_{k_2}(\eta')-\omega_{p_3}(\eta')-\omega_{p_4}(\eta') \big)\,d\eta' }
}{\Big[ \Omega_{k_1}(\eta_1)\Omega_{k_2}(\eta_1)\omega_{p_3}(\eta_1)\omega_{p_4}(\eta_1)\Big]^{1/2}} ~  \,,  \label{anniamp1} \ee  therefore, accounting for a symmetry factor $1/2 !$ for indistinguishable final states,  the self energy (\ref{sigmaA})  is given by
\be \Sigma[\eta_2;\eta_1] = \frac{\lambda^2}{8V} \int \frac{d^3p_3}{(2\pi)^3}\, \frac{e^{i \int^{\eta_2}_{\eta_1} \big(\Omega_{k_1}(\eta')+\Omega_{k_2}(\eta')-\omega_{p_3}(\eta')-\omega_{p_4}(\eta') \big) ~d\eta'}}{\Big[\mathcal{W}(\eta_1)\,\mathcal{W}(\eta_2)\Big]^{1/2}}  \,, \label{sigdef}\ee with
\be \mathcal{W}(\eta_1) = \Omega_{k_1}(\eta_1)\Omega_{k_2}(\eta_1)\omega_{p_3}(\eta_1)\omega_{p_4}(\eta_1)\,, \label{capW}\ee and $p_4 = |\vp_3-\vk_1-\vk_2|$.
Therefore the transition rate (\ref{rate}) is given by
\be \Gamma_{i\rightarrow f}(\eta) = 2 \int^{\eta}_{\eta_i} d\eta_1 \mathrm{Re}\Sigma[\eta;\eta_1]\,. \label{rate2}\ee

It is illuminating to relate the above results to the usual formulation of transition rates   from an initial state $\ket{i}$ at the initial time $\eta_i$ to a final state $\ket{f}$ at time $\eta$. The transition amplitude  is given by
\be \mathcal{A}_{i\rightarrow f}(\eta) = \bra{f}U_I(\eta,\eta_i)\ket{i} \label{amp} \ee and the transition  probability by
\be \mathcal{P}_{i\rightarrow f}(\eta) = |\bra{f}U_I(\eta,\eta_i)\ket{i}|^2 \,. \label{probi} \ee
To leading order in the coupling we find
\bea  \mathcal{A}_{i\rightarrow f}(\eta) &  =  & \frac{-i\,\lambda}{2\,V^2} \int d^3 x ~ e^{i\big(\vk_1+\vk_2 -\vp_3-\vp_4 \big)\cdot \vx }\int^{\eta}_{\eta_i}   ~~\frac{e^{-i \int^{\eta_1}_{\eta_i} \big(\Omega_{k_1}(\eta')+\Omega_{k_2}(\eta')-\omega_{p_3}(\eta')-\omega_{p_4}(\eta') \big) ~d\eta'}
}{\Big[ \Omega_{k_1}(\eta_1)\Omega_{k_2}(\eta_1)\omega_{p_3}(\eta_1)\omega_{p_4}(\eta_1)\Big]^{1/2}} ~  d\eta_1\nonumber \\ & = &   \frac{-i\,\lambda}{2\,V}\,\delta_{\vk_1+\vk_2 - \vp_3 -\vp_4}~~\int^{\eta}_{\eta_i}   ~~\frac{e^{-i \int^{\eta_1}_{\eta_i} \big(\Omega_{k_1}(\eta')+\Omega_{k_2}(\eta')-\omega_{p_3}(\eta')-\omega_{p_4}(\eta') \big) ~d\eta'}
}{\Big[ \Omega_{k_1}(\eta_1)\Omega_{k_2}(\eta_1)\omega_{p_3}(\eta_1)\omega_{p_4}(\eta_1)\Big]^{1/2}} ~  d\eta_1\,.  \label{anniamp} \eea  The total transition probability to the final states is given by
\be \mathcal{P}_{i\rightarrow f}(\eta) = \frac{1}{2!}\sum_{\vp_3}\sum_{\vp_4} |\mathcal{A}_{i\rightarrow f}(\eta)|^2\,, \label{PtotA} \ee which can be written compactly as
\be \mathcal{P}_{i\rightarrow f}(\eta) = \int^\eta_{\eta_i} d\eta_2 \int^\eta_{\eta_i} d\eta_1 \Sigma[\eta_2;\eta_1] \,,\label{probsig} \ee where $\Sigma[\eta_2;\eta_1]$ is given   by (\ref{sigdef}). Introducing $\Theta(\eta_2-\eta_1)+\Theta(\eta_1-\eta_2)=1$,   in (\ref{probsig}), where $\Theta$ is the Heaviside step function, relabelling $\eta_1 \leftrightarrow \eta_2$ in the term with $\Theta(\eta_1-\eta_2)$  and using that $\Sigma[\eta_1;\eta_2] = \Sigma^{*}[\eta_2;\eta_1]$ we find
\be \mathcal{P}_{i\rightarrow f}(\eta) =  2\,\int^\eta_{\eta_i} d\eta_2 \int^{\eta_2}_{\eta_i} d\eta_1 \mathrm{Re}\Sigma[\eta_2;\eta_1]\,, \label{ptoto}\ee which coincides with $\mathcal{P}_{tot}$ given by eqn. (\ref{Ptot}). Thus  {defining} the  comoving transition \emph{rate} as
\be \Gamma_{i\rightarrow f}(\eta) = \frac{ d }{d\eta} \mathcal{P}_{i\rightarrow f}(\eta) \label{sarate}\ee one finds the result (\ref{rate}) obtained from unitarity and the optical theorem. Consistently with the definition (\ref{rate}) of the transition rate we  {define} the  comoving   cross section as the transition rate per unit comoving flux for one incoming  particle, namely,
\be \sigma(\eta) = \frac{\Gamma_{i\rightarrow f}(\eta)}{\Big( \frac{v_{rel}(\eta)}{V}\Big)}\,, \label{Xsec}\ee with $v_{rel}(\eta)$ being the   comoving relative velocity, for collinear pair annihilation it  is given  by
\be v_{rel}(\eta) = \frac{|\vk_1|}{\Omega_{k_1}(\eta)}+\frac{|\vk_2|}{\Omega_{k_2}(\eta)} \,. \label{vrelpairann}\ee  We note that
\be \frac{|\vk|}{\Omega_k(\eta)} = \frac{|\vk_{ph}(\eta)|}{E^{\phi}_k(\eta)} \,.\label{physrat} \ee where
\be \vk_{ph}(\eta) = \frac{\vk}{a(\eta)}~~;~~ E^{\phi}_k(\eta) = \sqrt{k^2_{ph}(\eta)+M^2} = \frac{\Omega_k(\eta)}{a(\eta)} \,, \label{physkE}\ee  are the physical momenta and    energy   measured by a locally inertial observer. The time dependence of the relative velocity is a simple consequence of the cosmological redshift.

In section (\ref{sec:discussion}) we discuss subtleties and caveats emerging from the treatment during a finite time interval as is necessary within the cosmological setting.

\vspace{2mm}

\section{Massless $\chi$ particles:}\label{sec:massless} The integral over $\eta_1$ yielding $\Gamma_{i\rightarrow f}$ cannot be done analytically, even a numerical attempt is a daunting challenge because of the large range of $a(\eta)$ and momenta that
must be explored numerically. Instead, our strategy is to leverage the adiabatic approximation and the wide separation of time scales that it entails. We first study the case of \emph{massless} $\chi$- particles in the final state, the lessons of which will prove useful in the more general case of massive particles. For massless $\chi$ particles we find
\be 2 \,\Sigma[\eta;\eta_1] = \frac{\lambda^2}{4V}\, \frac{e^{i \int^{\eta}_{\eta_1} \big(\Omega_{k_1}(\eta')+\Omega_{k_2}(\eta')\big)d\eta'}}{\Big[\Omega_{k_1}(\eta)\Omega_{k_2}(\eta) \Omega_{k_1}(\eta_1)\Omega_{k_2}(\eta_1)\Big]^{1/2}} ~~ I[\eta;\eta_1] \,, \label{zeromasschi}\ee where we introduced the time kernel
\be I[\eta;\eta_1] =  {\int \frac{d^3p}{(2\pi)^3}\,\frac{e^{-i   \big(p+\big|\vK-\vp\big|\big)\,(\eta-\eta_1)}}{p\,\big|\vK-\vp\big|}} ~~;~~   \vK = \vk_1+\vk_2 \,. \label{masslesskernel}\ee   The momentum integral  can be done by introducing a convergence factor $\eta-\eta_1 \rightarrow \eta-\eta_1-i\epsilon$ with $\epsilon \rightarrow 0^+$, yielding
\be I[\eta;\eta_1] = -\frac{i}{4\pi^2} \, \frac{e^{-i|\vK|(\eta-\eta_1-i\epsilon)}
}{\eta-\eta_1-i\epsilon} = \frac{e^{-i|\vK|(\eta-\eta_1-i\epsilon)}}{4\pi^2} \,\Bigg\{-i\mathcal{P}\Big( \frac{1}{\eta-\eta_1} \Big) + \pi\,\delta(\eta-\eta_1) \Bigg\}\,,  \label{pint} \ee where $\mathcal{P}$ stands for the principal part. The ``short distance'' singularity as $\eta \rightarrow \eta_1$ is the same as in Minkowski space-time and stems from the large momentum behavior of the integral in $p$, namely linear in $p$. This linear divergence is manifest as the $\simeq 1/(\eta-\eta_1)$ as $\eta \rightarrow \eta_1$. Such ``short-distance'' singularity  remains even when the particles in the final  state are massive (see appendix (\ref{app:specdens})).

Introducing this result into equations (\ref{rate2},\ref{Xsec}) and gathering terms we obtain
\be \sigma(\eta) = \frac{\lambda^2}{16 \,\pi \Omega_{k_1}(\eta) \Omega_{k_2}(\eta) v_{rel}(\eta) } \, \frac{1}{2}\Bigg[ 1 + \frac{2}{\pi} \int^{\eta}_{\eta_i}\,\mathcal{R}_{12}[\eta;\eta_1]\, \frac{\sin[J(\eta;\eta_1)]}{\eta-\eta_1} \,  d\eta_1 \Bigg]\,, \label{Xsecfin}\ee
where
\be \mathcal{R}_{12}[\eta;\eta_1] = \Bigg[ \frac{\Omega_{k_1}(\eta) \Omega_{k_2}(\eta)}{\Omega_{k_1}(\eta_1) \Omega_{k_2}(\eta_1)} \Bigg]^{1/2} \,,\label{capR12}\ee and
\be J(\eta;\eta_1) = \int^{\eta}_{\eta_1}\, \Big(\Omega_{k_1}(\eta')+ \Omega_{k_2}(\eta')-\big| \vK \big| \Big) \,d\eta' \,. \label{bigJ}\ee The flux factor can be written in an illuminating manner,
namely
\be \Omega_{k_1}(\eta) \Omega_{k_2}(\eta) v_{rel}(\eta) = a^2(\eta)\, \Big[\big(\mathcal{P}_1(\eta)\cdot \mathcal{P}_2(\eta)\big)^2-M^4 \Big]^{1/2} \,,\label{fluxmink}\ee  and
\be \mathcal{P}_1(\eta)\cdot \mathcal{P}_2(\eta) = \eta_{ab} \, \mathcal{P}^{a}_1(\eta)\mathcal{P}^{b}_2(\eta) \label{minkflux} \ee
in terms of the Minkowski metric $\eta_{ab} = \mathrm{diag}(1,-1,-1,-1)$ and the four vectors in the local inertial frame
\be \mathcal{P}^{a}(\eta)= \Big(E^{\phi}_{k}(\eta),{\vec{k}}_{ph}(\eta)\Big)\,, \label{localk}\ee  where $\vk_{ph}(\eta), E_k(\eta)$ are given by eqn. (\ref{physkE}) for the respective particles. Therefore, up to the prefactor $a^2(\eta)$ the flux factor is the same as in Minkowski space-time but in terms of the local energies and momenta featuring the cosmological redshift, yielding
\be \sigma(\eta) = \frac{\lambda^2}{16 \,\pi \,a^2(\eta)\, \Big[\big(\mathcal{P}_1(\eta)\cdot \mathcal{P}_2(\eta)\big)^2-M^4 \Big]^{1/2} } \, \frac{1}{2}\Bigg[ 1 + \frac{2}{\pi} \int^{\eta}_{\eta_i}\,\mathcal{R}_{12}[\eta;\eta_1]\, \frac{\sin[J(\eta;\eta_1)]}{\eta-\eta_1} \,  d\eta_1 \Bigg]\,. \label{Xsecfin2}\ee The remaining time integral in (\ref{Xsecfin2}) cannot be done in closed form, nor is it useful to attempt a numerical study because of the large range of scales involved both in time and momenta. Instead, we implement the adiabatic approximation to extract its behavior and to be able to generalize to other processes. Using that during (RD) $a(\eta) = H_R\,\eta$, we  write:
\be \Omega^2_k(\eta_1) = k^2+M^2 a^2(\eta_1) = k^2+M^2 a^2(\eta)+ M^2 a^2(\eta)\Bigg[ \Big(\frac{\eta-\eta_1}{\eta}\Big)^2-2\,\Big(\frac{\eta-\eta_1}{\eta}\Big)\Bigg] \,, \label{adOme}\ee
we now introduce
\be  {\Omega_T}(\eta) = \Omega_{k_1}(\eta) + \Omega_{k_2}(\eta)\,,\label{totOme} \ee  as the total comoving energy scale of the process, and define:
\be \tau =  {\Omega_T}(\eta)(\eta-\eta_1)~~;~~ z(\eta) =  { {\Omega_T}(\eta)\,\eta}\,, \label{dimrats}\ee    In terms of these variables we obtain the relation
\be \Omega_k(\eta_1) =   \Omega_k(\eta)\, f_k(\tau) ~~;~~ f_k(\tau) = \Bigg[ 1 - \frac{2\, \tau}{\gamma^2_k(\eta)\,z(\eta)}\Big[1-\frac{ \tau}{2\, z(\eta)} \Big]\Bigg]^{1/2}   \,, \label{adOmefin}\ee where
\be \gamma_k(\eta) = \frac{\Omega_k(\eta)}{M\,a(\eta)} = \frac{E_k(\eta)}{M} \,, \label{lorfac}\ee is the local Lorentz factor. We note that
\be  z(\eta) = \frac{ E_{tot}(\eta)}{H(\eta)}  \,, \label{zeta}\ee  therefore,  the adiabaticity condition (\ref{adiarat}) implies that
\be   z(\eta) \gg 1 \,. \label{adzeta}\ee

The integral defining $J(\eta,\eta_1)$, eqn. (\ref{bigJ}), can be done explicitly by implementing the following steps: introducing
\be \Omega_T(\eta)\, \,\Big(\eta- \eta' \Big) \equiv x \,, \label{defx} \ee   it follows that
\be \Omega_k(\eta') = \Omega_k(\eta)\,f_k(x) \label{omeprim}\ee where $f_k(x)$ is given by eqn. (\ref{adOmefin}) with $\tau$ replaced by $x$. The integral (\ref{bigJ}) is now carried out in terms of the variable $x$ with the result
\be J(\eta,\eta_1) = J_0(\tau;\eta)+J_1(\tau;\eta)\,, \label{Jintfin}\ee where
\be J_0(\tau;\eta) =
\tau\,\Bigg[1- \frac{|\vec{K}|}{\Omega_T(\eta)}\Bigg]~~;~~ J_1(\tau;\eta) = \frac{\Omega_{k_1}(\eta)}{\Omega_T(\eta)}\,\Delta_{k_1}(\tau)+ \frac{\Omega_{k_2}(\eta)}{\Omega_T(\eta)}\,\Delta_{k_2}(\tau)   \,, \label{J0J1} \ee  where for each frequency ($1,2$)
\be \Delta(\tau) = \frac{\gamma z}{2} \Bigg\{ \big(\frac{1}{\gamma}-2\, \frac{\tau}{\gamma\,z}\big)- \big(\frac{1}{\gamma}- \,\frac{\tau}{\gamma\, z}\big)\, f_k(\tau) - \frac{\gamma^2  -1}{\gamma^2}\,\ln\Bigg[\frac{\big(\frac{1}{\gamma}- \frac{\tau}{\gamma\,z}\big)+ f_k(\tau)}{\frac{1}{\gamma}+1} \Bigg] \Bigg\} \simeq - \frac{ \tau^2}{2\,\gamma^2\,z}+ \cdots  \label{deltas} \ee  with $f_k(\tau)$   given by eqn. (\ref{adOmefin}) and we have suppressed the arguments $k;\eta$ in $\gamma;z$. For $\tau/\gamma z \ll 1$ it follows that
\be  \Delta(\tau) \simeq - \frac{ \tau^2}{2\,\gamma^2\,z}+ \cdots  \label{smaltaudel} \ee

The important aspect is that $J_0(\tau;\eta)$ is zeroth order adiabatic, whereas $J_1(\tau;\eta)$ is of first and higher adiabatic order, because $1/z = H(\eta)/E_{tot}(\eta) \ll 1$. The integral term in eqn. (\ref{Xsecfin2}) can now be written as
\be  \frac{2}{\pi} \int^{\eta}_{\eta_i}\,\mathcal{R}_{12}[\eta;\eta_1]\, \frac{\sin[J(\eta;\eta_1)]}{\eta-\eta_1} \,  d\eta_1  = \mathcal{S}_0(\eta)+\mathcal{S}_1(\eta) \ee where
\bea \mathcal{S}_0(\eta) & = & \frac{2}{\pi} \int^{\overline{z}(\eta)}_{0}\,\Bigg[\frac{1}{f_{k_1}(\tau)\,f_{k_2}(\tau)}\Bigg]^{1/2}\, \frac{\sin[J_0(\tau;\eta)]\cos[J_1(\tau;\eta)]}{\tau} \,  d\tau \,, \label{S0}\\
\mathcal{S}_1(\eta) & = & \frac{2}{\pi} \int^{\overline{z}(\eta)}_{0}\,\Bigg[\frac{1}{f_{k_1}(\tau)\,f_{k_2}(\tau)}\Bigg]^{1/2}\, \frac{\sin[J_1(\tau;\eta)]\cos[J_0(\tau;\eta)]}{\tau} \,  d\tau \,, \label{S1} \eea
and  we introduced
\be \overline{z}(\eta) = z(\eta)\,\Big(1-\frac{\eta_i}{\eta}\Big) ~~;~~ \frac{\eta_i}{\eta} \ll 1 \,.  \label{zbar}\ee

Since in the integrals $0\leq \tau \leq z$, for local Lorentz factors $\gamma \gg 1$ it  is clear from eqn. (\ref{smaltaudel}) that $J_1 \ll 1$ in the whole integration domain, therefore the contribution from $\mathcal{S}_0$ dominates
and $\mathcal{S}_1$ can be safely neglected, furthermore   in eqn. (\ref{S0}) we can replace the term $\cos[J_1] \simeq 1$. The case of non-relativistic particles with $\gamma \simeq 1$  but consistently with the adiabatic expansion  $z \gg 1$ requires further analysis since near the upper limit $\tau \simeq z$ and, in principle, the higher order adiabatic corrections may yield contributions comparable to the zeroth order adiabatic.

\begin{figure}[ht]
\begin{center}
\includegraphics[height=4in,width=4.5in,keepaspectratio=true]{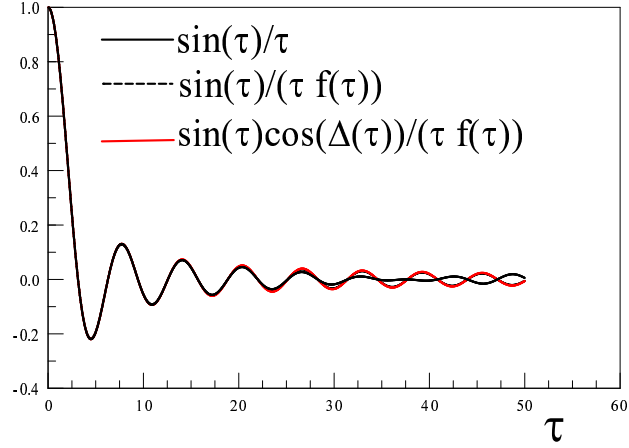}
\caption{ Integrand of $\mathcal{S}_0$ for $z = 100 ~,~\gamma=2$:  $\sin(\tau)/\tau$;
$\sin(\tau)/(\tau \,f(\tau))$;  $\sin(\tau)\,\cos(\Delta(\tau))/(\tau \,f(\tau))$ vs. $\tau$. The difference between the expressions is visible only at large $\tau$. }
\label{fig:integrand0}
\end{center}
\end{figure}

\begin{figure}[ht]
\begin{center}
\includegraphics[height=4in,width=4.5in,keepaspectratio=true]{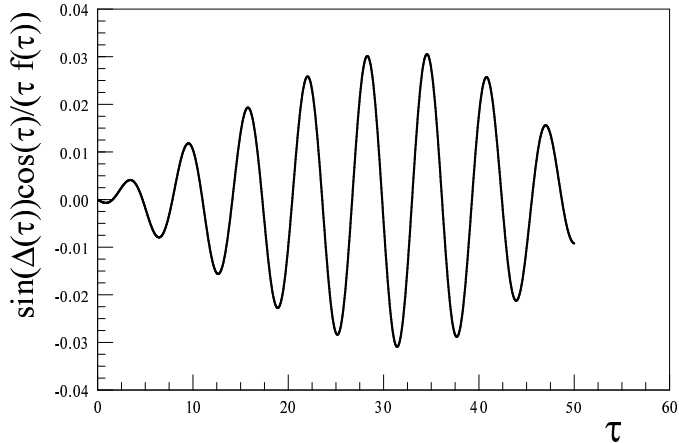}
\caption{ Integrand of $\mathcal{S}_1(\tau)$ vs $\tau$ for $z = 100;\gamma=2$.}
\label{fig:integrand1}
\end{center}
\end{figure}

The  contributions (\ref{S0},\ref{S1}) feature drastically different behavior: the integrand of $\mathcal{S}_0$ peaks at $\tau \simeq 0$ with an amplitude that is of $\mathcal{O}(1)$ and falls off very fast, whereas $\mathcal{S}_1$ vanishes $\propto \tau/\gamma\,z$ at small $\tau$,  is always of higher adiabatic order as it features powers of $1/z$, and oscillates rapidly, averaging out on long time scales. Furthermore, since the integrand  of   $\mathcal{S}_0$ is localized within the region $0\leq \tau \leq \pi$, it follows that in this region of integration  $\cos[J_1(\tau;\eta)]\simeq 1$ and $f_k(\tau) \simeq 1$ for $z\gg 1$. For $\tau \simeq z$ the   integrands  are of $\mathcal{O}(1/z) \ll 1$,   therefore the integrand of $\mathcal{S}_0$ can be replaced by $\sin[J_0(\tau;\eta)]$  and the contribution from $\mathcal{S}_1$ can be safely neglected for   $z \gg 1$. This analysis is verified numerically, figures (\ref{fig:integrand0}) and (\ref{fig:integrand1}) show  the integrands of $\mathcal{S}_0$ and $\mathcal{S}_1$  respectively for $\vk_1 = -\vk_2$ (center of mass) and $z = 100, \gamma=2$ with similar features for the full range of parameters with $z \gg 1$.

\begin{figure}[ht]
\begin{center}
\includegraphics[height=4in,width=4.5in,keepaspectratio=true]{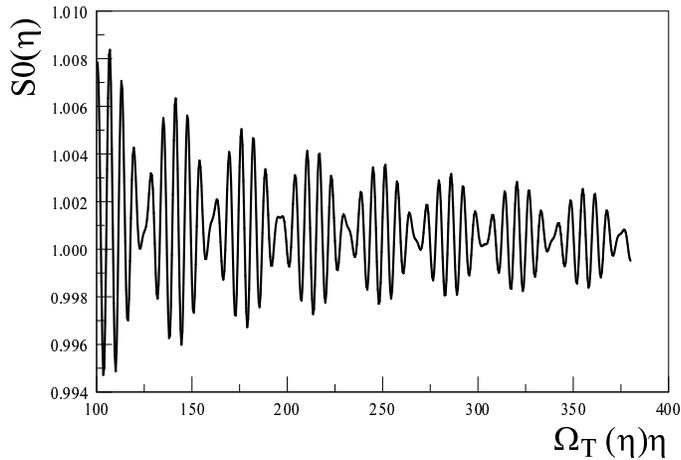}
\caption{  $\mathcal{S}_0(\eta)$ vs $\Omega_T(\eta)\eta$ for $\gamma=2$ and annihilation in the center of mass. For $\Omega_T(\eta)\eta \gg 1$ asymptotes to 1.}
\label{fig:szero}
\end{center}
\end{figure}

\begin{figure}[ht]
\begin{center}
\includegraphics[height=4in,width=4.5in,keepaspectratio=true]{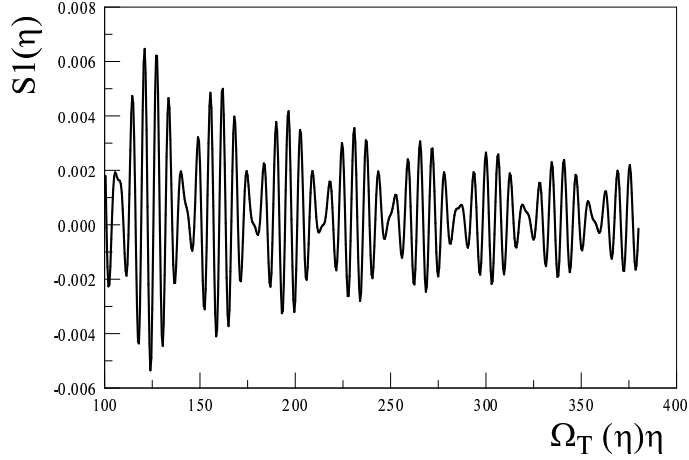}
\caption{  $\mathcal{S}_1(\eta)$ vs $\Omega_T(\eta)\eta$ for $\gamma=2$ and annihilation in the center of mass. For $\Omega_T(\eta)\eta \gg 1$ asymptotes to 0. Compare the vertical scale to that in fig. (\ref{fig:szero}). }
\label{fig:sone}
\end{center}
\end{figure}

The total integrals for $\mathcal{S}_0(\eta)$ and $\mathcal{S}_1(\eta)$ vs. $z(\eta)=\Omega_T(\eta) \eta $ are shown in figures (\ref{fig:szero}) and (\ref{fig:sone}) for pair annihilation in the center of mass for   $\gamma =2$.

\begin{figure}[ht]
\begin{center}
\includegraphics[height=4in,width=4.5in,keepaspectratio=true]{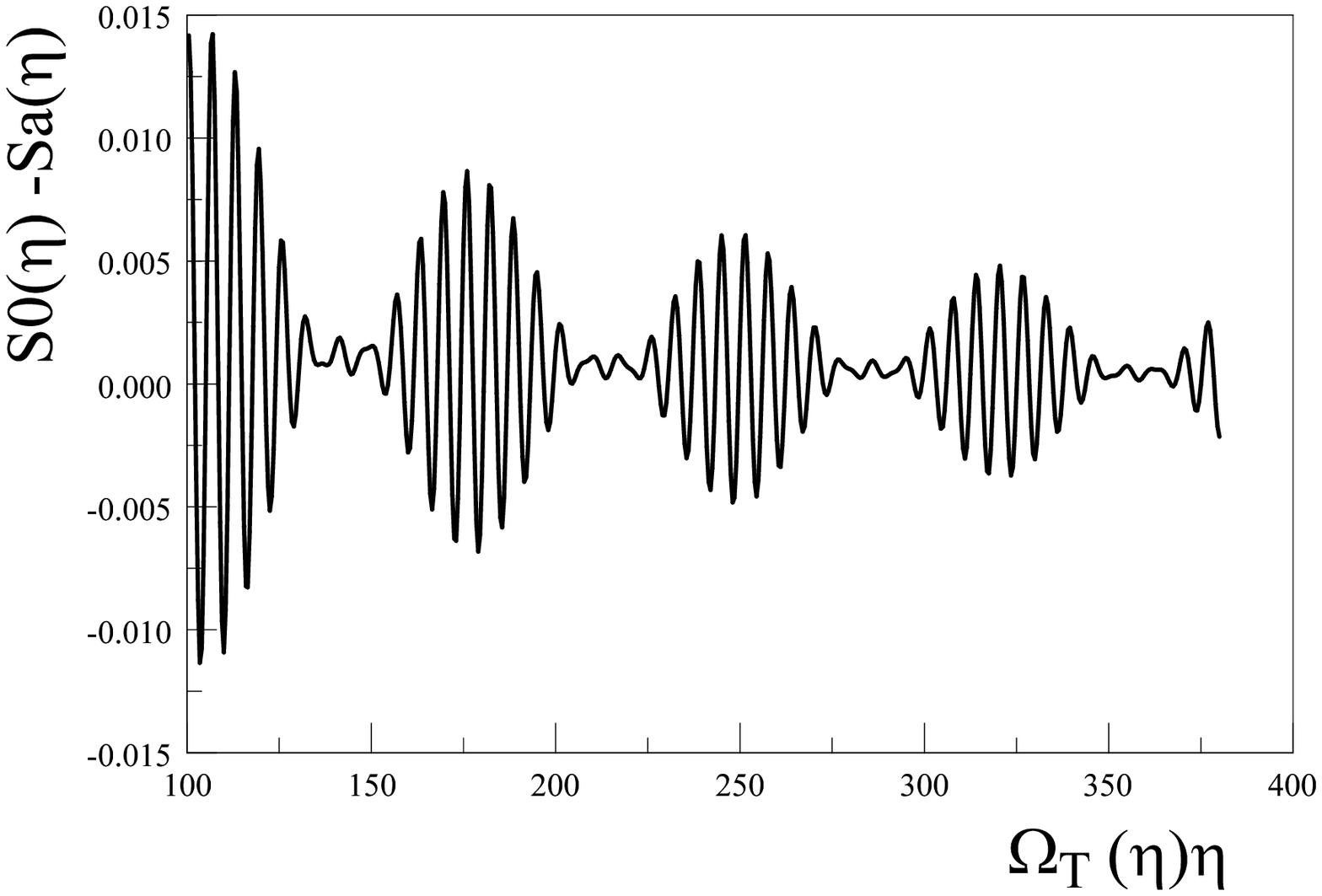}
\caption{  $\mathcal{S}_0(\eta)- S_a(\eta)$ with $S_a(\eta)=\frac{2}{\pi}Si[\Omega_T(\eta)\eta]$ vs $\Omega_T(\eta)\eta$ for $\gamma=2$ and annihilation in the center of mass.}
\label{fig:diference}
\end{center}
\end{figure}

Asymptotically for $z(\eta) = \Omega_T(\eta)\eta \gg 1$ the contributions $\mathcal{S}_0(\eta) \rightarrow 1$ and $\mathcal{S}_1(\eta) \rightarrow 0$. Since for large $z(\eta)= \Omega_T(\eta)\eta $ the terms $f_k(\eta) \rightarrow 1~~;~~\cos(\Delta(\eta)) \rightarrow 1$ and the integrand of $\mathcal{S}_0(\eta)$ is dominated by the region $\tau \simeq 0$,  for $\Omega_T(\eta)\eta \gg 1$ the contribution  $\mathcal{S}_0(\eta)$ can be well approximated by setting $f_k = 1~;~\cos(\Delta) =1$, in which case, for pair annihilation in the center of mass,  we can replace
\be \mathcal{S}_0(\eta) =   \frac{2}{\pi} Si[\Omega_T(\eta)\eta] \label{Sap}\ee where $Si[x]$ is the sine-integral function. Fig. (\ref{fig:diference}) displays $\mathcal{S}_0(\eta)-\frac{2}{\pi}Si[\Omega_T(\eta)\eta]$ vs $\Omega_T(\eta)\eta$ for $\gamma=2$ and annihilation in the center of mass showing that this difference becomes negligibly small  for $\Omega_T(\eta)\eta \gg 1$.

A main conclusion of this analysis, confirmed by the numerical study,  is that  the contribution from the two-body phase space integral  yields a \emph{time kernel} proportional to  $1/\tau$. It is the rapid fall off  of this kernel that ensures that the contributions of the adiabatic corrections $\propto \tau/z$ are \emph{suppressed} by $\mathcal{O}(1/z)$ as compared to the zeroth order terms. At the time scale $\tau \simeq z$ when  the terms of higher adiabatic order begin to be of the same order as the leading terms, the time kernel has been suppressed by $\simeq 1/z$ thereby suppressing their contributions. This is an important corollary of the simpler, massless case, it is the rapid fall-off of the time kernel that ensures the reliability of the adiabatic expansion in the time integrals, a result that is not obvious  \emph{a priori}.

 As we will see below,   this result holds generally for any mass of the outgoing particles with a time kernel that falls off \emph{faster} than $1/\tau$ for massive particles in the out state, thereby improving the reliability of the adiabatic expansion in the time integrals for the cross section.

This analysis also shows that in the  limit $z(\eta) = \Omega_T(\eta) \,\eta \gg 1$  the contribution from $\mathcal{S}_1(\eta)$ can be safely neglected. Similar results and conclusions are obtained for $\vk_1+\vk_2 \neq 0$ in which case, taking $\eta_i \rightarrow 0$
\be \mathcal{S}_0(\eta) = \frac{2}{\pi}\,Si[(\Omega_T(\eta) - K)\eta] ~~;~~ K = |\vk_1+\vk_2| \,. \label{noCM}\ee Gathering all these results, we find the final form of the pair annihilation cross section valid for $\Omega_T(\eta)\eta = E_T(t)/H(t) \gg 1$, where $\Omega_T(\eta)$ is defined in eqn. (\ref{totOme}) and $E_T(t)=\Omega_T(\eta)/a(\eta)$,

  \be \sigma(\eta) = \frac{\lambda^2}{16 \,\pi \,a^2(\eta)\, \Big[\big(\mathcal{P}_1(\eta)\cdot \mathcal{P}_2(\eta)\big)^2-M^4 \Big]^{1/2} } \, \frac{1}{2}\Big[ 1 + \frac{2}{\pi} Si\big[\big(\Omega_T(\eta) - K\big)\eta\big] \Big]\,. \label{Xsecfinasy}\ee Up to the scale factor dependence in the denominator, this result is remarkably similar to the cross section during a finite time interval in Minkowski space-time found in appendix (\ref{app:Xminkfiti}).
  Equations (\ref{sigminkfina},\ref{sigminkfina}) in the appendix should be compared to the result (\ref{Xsecfinasy}).

  The $1/a^2(\eta)$ dependence of the cross section has a simple interpretation: we have obtained a \emph{comoving} cross section, by obtaining the comoving transition rate and dividing by the comoving flux. Since the cross section has dimensions of area, upon cosmological expansion all physical lengths scale with $a(\eta)$ therefore, a \emph{physical} cross section should be identified with
   \be \sigma_{ph}(\eta)= a^2(\eta)\sigma(\eta)\label{sigmaphys} \,,\ee
    which in terms of the physical (local) four momenta and comoving time agrees with the cross section in Minkowski space-time during a finite time interval as shown explicitly in  appendix \ref{app:Xminkfiti} (see eqn. (\ref{sigminkfina})).

  The bracket in (\ref{Xsecfinasy}) has an important interpretation that paves the way towards understanding the general case of massive particles in the final state. This interpretation begins with writing the time kernel $I[\eta;\eta_1]$ in eqn. (\ref{masslesskernel}) in terms of a spectral representation, namely
  \be I[\eta;\eta_1] = \int^{\infty}_{-\infty} \rho(K_0,K)\,e^{-iK_0(\eta-\eta_1)} \,dK_0 \,, \label{Irhorep}\ee
  with
  \be \rho(K_0,K) = \int \frac{d^3p}{(2\pi)^3}\,\frac{\delta\big(K_0 - p-\big|\vK-\vp\big|\big)}{p \,\big|\vK-\vp\big|}  = \frac{1}{4\pi^2} \,\Theta(K^2_0-K^2)\,\Theta(K_0)  \,. \label{rhofrw} \ee The spectral density $\rho(K_0,K)$ is identified as the   Lorentz invariant phase space for two massless particles in Minkowski space-time.

    The result (\ref{pint}) and analysis above showed that the time kernel is dominated by the region  $0 \leq \tau \simeq  \pi$ and its rapid fall-off $\propto 1/\tau$ suppresses the integration region for $\tau \simeq z(\eta) \gg 1$. Therefore the integral $J(\tau;\eta)$ (eqn. (\ref{bigJ})) can be safely replaced by the zeroth-adiabatic order result $J_0(\tau;\eta)$ in eqn. (\ref{J0J1}), effectively replacing
    \be  e^{i \int^{\eta}_{\eta_1} \big(\Omega_{k_1}(\eta')+\Omega_{k_2}(\eta')\big)d\eta'} ~ \rightarrow ~ e^{i  \big[\Omega_{k_1}(\eta)+\Omega_{k_2}(\eta)\big](\eta-\eta_1)} \,, \label{exprepla}\ee   hence    neglecting the higher adiabatic order contribution $J_1(\tau;\eta)$, and the factor $\mathcal{R}_{12}[\eta;\eta_1]$ in eqn. (\ref{Xsecfin}) can be set to  $\mathcal{R}_{12}[\eta;\eta_1]=1$ to leading adiabatic order. Therefore, to leading (zeroth) adiabatic order, and now taking $\eta_i \rightarrow 0$,  we find the comoving transition rate $\Gamma_{i\rightarrow f}$ given by eqn. (\ref{rate2}) as
  \be \Gamma_{i\rightarrow f}(\eta) =  \frac{\lambda^2}{4V}\, \frac{1}{ \Omega_{k_1}(\eta)\Omega_{k_2}(\eta)}\, \int^\infty_{-\infty} \rho(K_0,K)\, \frac{\sin\Big[ \Big( K_0 - \Omega_T(\eta)\Big) \eta \Big]}{\Big( K_0 - \Omega_T(\eta)\Big)}\,dK_0  \,. \label{simplegama}\ee

  Although the transition rate features oscillations, under the same approximations keeping the leading (zeroth) adiabatic order, we find the integral over a short time interval during which the time dependent frequencies do not change much
  \be  \int^{\eta}_{0} \Gamma_{i\rightarrow f}(\eta') \, d\eta'= \frac{\lambda^2}{2V}\, \frac{1}{ \Omega_{k_1}(\eta)\Omega_{k_2}(\eta)}\, \int^\infty_{-\infty} \rho(K_0,K)\, \Bigg[\frac{\sin\Big[ \Big( K_0 - \Omega_T(\eta)\Big) \eta/2)\Big]}{\Big( K_0 - \Omega_T(\eta)\Big)}\Bigg]^2\,dK_0 \,.\label{gamainteg} \ee This result is manifestly positive as anticipated by the relation (\ref{Ptot}), and is reminiscent of Fermi's Golden rule:  formally,  the limit $\eta\rightarrow \infty$ inside the bracket in the integral in (\ref{gamainteg}) yields $\pi \,\eta\, \delta(K_0-\Omega_T(\eta))$, leading to  the usual result as in Minkowski space-time, as shown in appendix (\ref{app:Xminkfiti}) (see equation (\ref{interatemink})).
  However, taking the infinite time limit is clearly inconsistent with the time dependence of the (conformal) energies and the redshift of physical momenta.

  Using the result (\ref{rhofrw}) for the spectral density, changing
  variables to $(K_0-\Omega_T(\eta))(\eta-\eta_i)\equiv X$  and taking $\eta_i \rightarrow 0$,  the integral in (\ref{simplegama}) becomes
  \bea \int^\infty_{-\infty} \rho(K_0,K)\, \frac{\sin\Big[ \Big( K_0 - \Omega_T(\eta)\Big)\, \eta \Big]}{\Big( K_0 - \Omega_T(\eta)\Big)}\,dK_0 &  = &  \frac{1}{4\pi^2}\int^\infty_{-(\Omega_T(\eta)-K)\eta}   \frac{\sin\big[ X\big]}{X}\,dX \nonumber \\ & = &  \frac{1}{8\pi}\Bigg[1 + \frac{2}{\pi}\,Si\Big[\Big(\Omega_T(\eta)-K\Big)\eta \Big] \Bigg]\,,\label{finainteg}\eea  yielding  exactly the cross section (\ref{Xsecfinasy}).
  This analysis will prove useful to study the general case with massive particles in the initial and final states in the next section.

  In the limit $\big(\Omega_T(\eta) - K\big)\eta \rightarrow \infty$ the cross section  (\ref{Xsecfinasy}) becomes
  \be \sigma(\eta) = \frac{\lambda^2}{16 \,\pi \,a^2(\eta)\, \Big[\big(\mathcal{P}_1(\eta)\cdot \mathcal{P}_2(\eta)\big)^2-M^4 \Big]^{1/2} } \,, \label{Xsecfinity}\ee which is
  \emph{invariant under local Lorentz transformations} at a fixed conformal time, namely
  \be \mathcal{P}^a(\eta) \rightarrow \Lambda^a_b \mathcal{P}^b(\eta)\label{LLT}\ee where $\Lambda^a_b$ are the Lorentz transformation matrices at a fixed (conformal) time $\eta$.   However, for finite $\eta$, the $Si$ function  and consequently the cross section (\ref{Xsecfinasy}) is not invariant under the local Lorentz transformation.  A similar behavior is found for the cross section during finite time in Minkowski space-time derived in appendix (\ref{app:Xminkfiti}), during a finite time interval the cross section is \emph{not} Lorentz invariant (as expected) but Lorentz invariance is restored in the infinite time limit. Since $\eta$ is the particle horizon (for $\eta_i \rightarrow 0$), it follows that the finiteness of the particle horizon entails a violation of local Lorentz invariance. This important aspect is general   as discussed in the next section.

\vspace{1mm}

\textbf{Ultrarelativistic limit:} In the ultrarelativistic limit $\Omega_{k_{1,2}} \simeq k_{1,2}$
\be \sigma(\eta) = \frac{\lambda^2}{32\,\pi \,k_1\,k_2}\,\frac{1}{2}\Bigg[ 1 + \frac{2}{\pi} Si\big[\big(k_1+k_2 - K\big)\eta\big] \Bigg] \,, \label{URlim}\ee which is the same as for Minkowski space-time with the replacement $\eta \rightarrow t$. This behavior is expected since in the ultrarelativistic limit in conformal time in a (RD) cosmology the mode functions are exactly the same as in Minkowski space-time since the frequencies do not depend on time. This is   a manifestation of the equivalence principle.

\vspace{1mm}

\textbf{Non-relativistic limit:} in this limit $\Omega_k \simeq M\,a(\eta)$ and we find
\be \sigma(\eta) = \frac{\lambda^2}{16\,\pi \,M\, a(\eta)(k_1+k_2)}\,\frac{1}{2}\Bigg[ 1 + \frac{2}{\pi} Si\big[\big(2\,M\,a(\eta) - K\big)\eta\big] \Bigg] \,. \label{NRlim}\ee Taking the asymptotic limit so that $Si\big[\big(2\,M\,a(\eta) - K\big)\eta\big] \rightarrow \pi/2$ it follows that the physical cross section $a^2(\eta)\,\sigma(\eta)$  features the same form as in Minkowski space-time but in terms of the physical, redshifted momenta.

\section{General case: massive particles.}\label{sec:massive}
For the general case with massive particles in the final state, the time kernel introduced in eqns. (\ref{zeromasschi},\ref{masslesskernel}) now becomes
\be I[\eta;\eta_1] = \int \frac{d^3p}{(2\pi)^3} \frac{e^{-i \int^\eta_{\eta_1} \big(\omega^{(3)}_{p}(\eta')+\omega^{(4)}_{q}(\eta')\big)\,d\eta'}}{\Big[\omega^{(3)}_{p}(\eta)\,\omega^{(4)}_{q}(\eta)\Big]^{1/2}\Big[\omega^{(3)}_{p}(\eta_1)\,\omega^{(4)}_{q}(\eta_1)\Big]^{1/2}}~~;~~
\vec{q} = \vec{p}-\vec{K} \,, \label{massivekernel}\ee
with
\be \omega^{(i)}_{k}(\eta) = \sqrt{k^2+m^2_i\,a^2(\eta)} ~~;~~ i= 3,4 \label{outfreqsmass}\ee allowing the final state to be that of two particles of different masses as a general case.

Unlike the case of massless particles in the final state, we cannot perform the momentum integral in eqn. (\ref{massivekernel}), neither the time integral in eqn. (\ref{rate2}) in closed form.

To make progress we invoke the following lessons learned from the analysis of the previous section and that of  appendices (\ref{app:Xminkfiti},\ref{app:specdens}) wherein we study the cross section in real time and the time kernel in Minkowski space-time as a guide.

The study of the massless case in the previous section showed that the time kernel is localized at $\tau =\Omega_T(\eta -\eta_1) \lesssim 1$ and its rapid fall off $\propto 1/\tau$ for large $\tau$ suppresses the higher order adiabatic corrections $\propto \tau/z$. The behavior $1/\tau$ as $\tau \rightarrow 0$ is a consequence of the large momentum dominance of the integral in the time kernel. This is manifest in the spectral representation, eqn. (\ref{Irhorep}) where the spectral density $\rho(K_0,K) \rightarrow \mathrm{constant}$ as $K_0 \rightarrow \infty$. As shown in appendix (\ref{app:specdens}) the short time behavior is the same for massive or massless final states. Following the same steps leading up to eqn. (\ref{adOmefin}) for the frequencies of the incoming states, and in terms of $x=\Omega_T(\eta)(\eta-\eta')$ (see eqn. (\ref{defx})) and $z = \Omega_T(\eta)\eta$
we find
\be \omega^{(i)}_{k}(\eta') =   \omega^{(i)}_{k}(\eta)\, f^{(i)}_k(x) ~~;~~ i=3,4  \label{outfreqs}\ee
where
\be f^{(i)}_{k}(x) = \Bigg[1-g^{(i)}_{k}\Big[\frac{x}{z}\Big]\Bigg]^{1/2}~~;~~  g^{(i)}_{k}\big[w\big]=\frac{2\,w}{(\gamma^{(i)}_{k}(\eta))^2}\,\Big(1-\frac{w}{2}\Big)~~;~~ 0\leq w < 1 \,, \label{gofx}\ee and
$\gamma^{(i)}_k(\eta) =  {\omega^{(i)}_k(\eta)}/{m^{(i)}\,a(\eta)}  $ for each species.
The function $g^{(i)}_k[x/z]$ encodes the corrections to the leading adiabatic order as is explicit in the $1/z$ factor. It attains its maximum for $\gamma^{(i)}_{k}=1$, which yields  the largest deviation from
the adiabatic zeroth order, and in this case,
\begin{equation}
f^{(i)}_{k}(x)=1-x \,,\label{ub}
\end{equation}
  yielding an \emph{upper bound} on the corrections to the leading adiabatic order. We now replace this upper bound into (\ref{massivekernel}) obtaining
  \begin{equation}
I^{(ub)}[\eta,\eta_1]=\frac{1}{(1-\frac{\tau}{z})}\int\frac{d^{3}p}{(2\pi)^{3}}
\frac{e^{-i(\omega^{(3)}_{p}(\eta)+\omega^{(4)}_{q} (\eta))T}}{\omega_{p}^{(3)}(\eta)~\omega_{q}^{(4)}(\eta)}~~;~~ \vec{q}=\vec{p}-\vec{K}\,,
\label{Iub}
\end{equation} where we defined
\begin{equation}
T=(\eta-\eta_1)\Big[1-\frac{\tau}{2z}\Big]  \,. \label{capT}
\end{equation}
The superscript $(ub)$ in (\ref{Iub}) refers to the fact that this time kernel gives an upper bound to the corrections beyond the leading adiabatic order. Under this upper-bound approximation we can now cast the momentum integral as a spectral representation just as in the case of Minkowski space-time in appendix (\ref{app:specdens}), namely
\be \int\frac{d^{3}p}{(2\pi)^{3}}
\frac{e^{-i(\omega^{(3)}_{p}(\eta)+\omega^{(4)}_{q} (\eta))T}}{\omega_{p}^{(3)}(\eta)~\omega_{q}^{(4)}(\eta)} = \int^{\infty}_{-\infty} \rho(K_0,K) e^{-iK_0\,T} \,dK_0 \,, \label{specrep}\ee with
\be \rho(K_0,K) = \int\frac{d^{3}p}{(2\pi)^{3}}
\frac{\delta\Big(K_0-\omega_{p}^{(3)}(\eta)-\omega_{p}^{(4)}(\eta)\Big)}{\omega_{p}^{(3)}(\eta)~\omega_{q}^{(4)}(\eta)}
\,,\label{nurho}\ee being exactly the spectral density in Minkowski space-time analyzed in detail in appendix (\ref{app:specdens}) but depending parametrically on $\eta$ and  given by
\be \rho(K_0,K) = \frac{1}{4\pi^2} \Bigg[1- \frac{(m_3-m_4)^2\,a^2(\eta)}{K^2_0-K^2} \Bigg]^{1/2}\,\Bigg[1- \frac{(m_3+m_4)^2\,a^2(\eta)}{K^2_0-K^2} \Bigg]^{1/2} \,\Theta\big(K_0-K_T(\eta)\big) \, , \label{nu2prho}\ee where the
threshold $K_T(\eta)$ is
\be K_T (\eta)= \sqrt{K^2+(m_3+m_4)^2\,a^2(\eta)} ~~;~~ \vec{K} = \vec{k_1}+\vec{k_2} \,. \label{nuthreshold} \ee We can now use the results established in appendix (\ref{app:specdens}) to extract the short and long time behavior of the upper bound time kernel given by eqn. (\ref{Iub}).

\vspace{1mm}

\textbf{i:) Short time regime.} This region   corresponds to  $K_T\, T\ll 1$, in this case $\eta_1 \simeq \eta$, therefore $\tau/z \ll 1$.  The short time behavior is the same for massless or massive particles in the final state and yields
\be I^{(ub)} \simeq
 -\frac{i}{4\pi^{2}}\,\frac{e^{-i\,K_{T}\,(\eta-\eta_1)}}{ (\eta-\eta_1)-i\epsilon }+\mathrm{finite} \,. \label{stIub}
\end{equation}  where \emph{finite} stands for a finite constant as $\eta \rightarrow \eta_1$, the short time behavior (\ref{stIub})  is similar to the result (\ref{pint}). In this short time region, the corrections $\propto \tau/z$ to the leading adiabatic order can be neglected, as analyzed in detail in the previous section for massless particles in the final state.

\vspace{1mm}

\textbf{ii:) Long  time regime: $K_T\,T \gg 1$.} In appendix (\ref{app:specdens}) we find that if the spectral density vanishes at threshold as $\rho(K_0,K) \propto (K_0-K_T)^\alpha$ as $K_0 \rightarrow K_T$, the   long time behavior of the time kernel (\ref{specrep}) is $\propto 1/K^\alpha_T T^{\alpha+1}$ (see eqn. (\ref{asygeneral})). For both massless particles in the final state $\alpha=0$ yielding the behavior $1/T$, for one massless and one massive particle in the final state $\alpha=1$ and the long time behavior is $\propto 1/T^2$ and for both massive particles $\alpha = 1/2$ yielding the long time behavior $1/T^{3/2}$. In order to analyze the contribution from the corrections to the zeroth adiabatic order terms, it is convenient to analyze the short and long time behavior in terms of the variables $z= \Omega_T(\eta)\eta$ and  $\tau = \Omega_T(\eta)(\eta-\eta_1)$ introduced in the previous section. In these variables the time integrals restrict $\tau$ to the interval $0\leq \tau \leq \overline{z}$ with $\overline{z}$ given by eqn. (\ref{zbar}) and $\eta_i/\eta \ll 1$,    as in the massless case of the previous section. The short time region corresponds to $\tau \ll z$ during which the corrections to the zeroth order adiabatic $\propto \tau/z$ are negligible, as discussed in the massless case. For the case of massive particles in the final state, the long time behavior of the upper bound approximation to the time kernel $I^{(ub)}$ (\ref{Iub}) yields for $\alpha = 1/2,1$

\be I^{(ub)}[\tau,z] \propto \frac{1}{\tau^{\alpha}\,\big(1-\frac{\tau}{2z}\big)^{1+\alpha}} \Bigg[\frac{1}{\tau} + \frac{1}{z-\tau} \Bigg]\,. \label{Iublongtime}\ee  Near the end point of the time integrals $\tau \simeq \overline{z}$ the second term in the bracket in (\ref{Iublongtime}) dominates. From the definition of $\overline{z}$, eqn. (\ref{zbar})  it follows that
\be z-\overline{z} = z\,\frac{\eta_i}{\eta}= \Omega_T(\eta) \eta_i = \frac{\Omega_T(\eta)}{\Omega_T(\eta_i)}\, z_i\,, \label{zdif}\ee  where
\be z_i = \Omega_T(\eta_i)\eta_i \gg 1\,, \label{zini}\ee because for consistency  the adiabatic approximation must hold at the initial time. Furthermore, since $\Omega_T(\eta)/\Omega_T(\eta_i) \geq 1$ it follows that near the upper limit of integration
\be I^{(ub)}[\tau \simeq \overline{z},z] \lesssim \frac{1}{z_i\,z^{\alpha}} \ll 1 \,. \label{ubIub} \ee Hence,  the  integration region where the corrections to the zeroth adiabatic order become important, namely $\tau \simeq \overline{z}$    is strongly suppressed by the rapid fall-off   of the time kernel by a factor $1/(z_i z^\alpha) \ll 1$. Furthermore, since $I^{(ub)}$ is an upper bound, the contribution  of this region of integration is even smaller than the bound from eqn. (\ref{ubIub}).
This analysis is valid  for any masses of final state particles and confirms that the time kernel strongly suppresses the region of integration   $\tau \simeq z$ in which the higher order adiabatic corrections could compete with the zeroth order. This result is in   agreement with the case of massless final states studied in the previous section, and shows that the case of massive particles in the final state is even more suppressed than the massless one. This analysis demonstrates that  we can safely neglect the higher order adiabatic corrections encoded in the contributions $\propto \tau/z,\tau^2/z^2 \cdots$,   keeping solely the zeroth order contributions. This  is tantamount to replacing $f^{(i)}_k(x) \rightarrow 1$ in all frequencies associated with the final states (see eqns. (\ref{outfreqs},\ref{gofx})), as well as $f_k(\tau) \rightarrow 1$ in all frequencies of the initial states (see eqn. (\ref{adOmefin})). These replacements yield the   time kernel (\ref{massivekernel})
\be I[\eta;\eta_1] = \int \frac{d^3p}{(2\pi)^3} \frac{e^{-i   \big(\omega^{(3)}_{p}(\eta)+\omega^{(4)}_{q}(\eta)\big)(\eta-\eta_1)}}{\omega^{(3)}_{p}(\eta)\,\omega^{(4)}_{q}(\eta)}
 = \int^{\infty}_{-\infty} \rho(K_0,K) e^{-iK_0(\eta-\eta_1)} \,dK_0 \,, \label{massivekernelzero}\ee where the spectral density $\rho(K_0,k)$ is given by (\ref{nu2prho}). Incorporating these replacements in eqn. (\ref{zeromasschi}) yields
\be 2 \,\Sigma[\eta;\eta_1] = \frac{\lambda^2}{4V}\, \frac{e^{i   \Omega_T(\eta)(\eta-\eta_1)}}{ \Omega_{k_1}(\eta)\Omega_{k_2}(\eta) } ~~ I[\eta;\eta_1] \,, \label{zerozeromasschi}\ee with $\Omega_T(\eta)= \Omega_{k_1}(\eta)+\Omega_{k_2}(\eta)$. We now   obtain the transition rate $\Gamma_{i\rightarrow f}$, eqn. (\ref{rate2}) by carrying out the \emph{time integral} in $\eta_1$, and from eqn. (\ref{Xsec}) the cross section to leading order in the adiabatic approximation for particles in the initial state with arbitrary masses $M_{1,2}$ is given by
\be \sigma(\eta) = \frac{\lambda^2}{4 \,a^2(\eta)\, \Big[\big(\mathcal{P}_1(\eta)\cdot \mathcal{P}_2(\eta)\big)^2-M^2_1\,M^2_2 \Big]^{1/2} } \, \int^{\infty}_{-\infty} \rho(K_0;K) \, \frac{\sin\Big[\big(K_0 -\Omega_T(\eta)\big)(\eta-\eta_i) \Big]}{\big(K_0 -\Omega_T(\eta)\big)} \,dK_0 \,,\label{FGR}\ee where we have used the relation to the flux given by (\ref{fluxmink}). For massive particles with masses $m_{3},m_{4}$ in the final state the spectral density is given by eqns. (\ref{nu2prho},\ref{nuthreshold}).

 This is our main result for the general case of massive particles both in the initial and final state with masses $M_{1,2}$ and $m_{3,4}$ respectively. It is now straightforward to confirm that for the case of massless particles in the final state, the spectral density is given by (\ref{rhofrw}) and using the result given by eqn. (\ref{finainteg}) the (pair annihilation) cross section is
given precisely by eqn. (\ref{Xsecfinasy}).

\subsection{Pair annihilation:  $\overline{\phi}\phi \rightarrow \chi \chi$}\label{subsec:pair}

In this case the initial state particles are of equal mass $M_1=M_2=M$ and final state particles are also of equal mass $m_3=m_4=m \neq 0$. We take $\eta_i \rightarrow 0$ and changing variables to $K_0 = \Omega_T(\eta)+ X/\eta$ we find
\be \sigma(\eta) = \frac{\lambda^2}{16\,\pi^2 \,a^2(\eta)\, \Big[\big(\mathcal{P}_1(\eta)\cdot \mathcal{P}_2(\eta)\big)^2-M^4 \Big]^{1/2} } \, \int^{\infty}_{-\overline{X}(\eta)} \widetilde{\rho}( X;\eta) \,\, \frac{\sin\big[X\big]}{X} \,dX \,,\label{sigmasfin} \ee where
\be \overline{X}(\eta) =  \frac{ E_T(\eta) - K_{Tp}(\eta) }{H(\eta)} ~~;~~E_T(\eta) = \frac{\Omega_T(\eta)}{a(\eta)}~~;~~ K_{Tp}(\eta)=\Big[K^2_{ph}(\eta)+ 4\,m^2 \Big]^{1/2}~~;~~ K_{ph}(\eta) = \frac{K}{a(\eta)}\,, \label{Xmin}\ee with $H(\eta) = a'(\eta)/a^2(\eta)= 1/(a(\eta)\eta)$ the Hubble expansion rate in (RD), and
\be \widetilde{\rho}( X;\eta) = \Bigg[1- \frac{4m^2}{\mathcal{S}(\eta)+X^2 H^2(\eta)+2XE_T(\eta)H(\eta)}   \Bigg]^{\frac{1}{2}}\,, \label{rotil}\ee where
\be \mathcal{S}(\eta) = \Big( \mathcal{P}_1(\eta)+\mathcal{P}_2(\eta)\Big)^2 = E^2_{T}(\eta) - K^2_{ph}(\eta) \,,\label{smandel}\ee is the local Mandelstam variable depending adiabatically on time through the local four momenta (\ref{localk}) with the Minkowski scalar product (\ref{minkflux} ), therefore $\mathcal{S}(\eta)$     is   invariant under the local Lorentz transformations (\ref{LLT}). However, for $H(\eta) \neq 0$ the cross section (\ref{sigmasfin}) is not invariant under local Lorentz transformations: both the spectral density which depends explicitly on $E_T$ and  the lower limit $\overline{X}$ which depends both on $E_T$ and $K_{Tp}$ violate explicitly the invariance under local Lorentz transformations. We note that the physical cross section $\sigma_{ph}(\eta) = a^2(\eta)\,\sigma(\eta)$ with $\sigma(\eta)$ given by (\ref{sigmasfin}) is strikingly similar to the cross section in Minkowski space-time but during a finite time interval, given by eqn.  (\ref{sigrhot}) with the spectral density (\ref{rotil}) with the replacement $H(\eta) \rightarrow 1/t$.

The infinite time limit corresponds to $\eta\rightarrow \infty$, namely $H(\eta) \rightarrow 0$. In this limit   the $X$ integral in eqn. (\ref{sigmasfin}) yields
\be \int^{\infty}_{-\overline{X}(\eta)} \widetilde{\rho}( X;\eta) \,\, \frac{\sin\big[X\big]}{X} \,dX~~~  {}_{\overrightarrow{\eta \rightarrow \infty}} ~~ \pi \, \Bigg[1- \frac{4m^2}{\mathcal{S}(\eta)}   \Bigg]^{\frac{1}{2}}\, \Theta\Big(\mathcal{S}(\eta)-4m^2\Big) \,. \label{inflimi} \ee In this ``infinite time limit'' the cross section becomes
\be \sigma(\eta) ~~{}_{\overrightarrow{\eta \rightarrow \infty}}~~\frac{\lambda^2\,\Big[1- \frac{4m^2}{\mathcal{S}(\eta)}   \Big]^{\frac{1}{2}}}{16\,\pi  \,a^2(\eta)\, \Big[\big(\mathcal{P}_1(\eta)\cdot \mathcal{P}_2(\eta)\big)^2-M^4 \Big]^{1/2} }\, \Theta\Big(\mathcal{S}(\eta)-4m^2\Big)\,, \label{siginfi} \ee where   the $\Theta\Big(\mathcal{S}(\eta)-4m^2\Big)$ function reflects the kinematic threshold depending adiabatically on time through the red-shifted momenta. Up to the explicit dependence on the scale factor in the denominator, at a fixed $\eta$ this is the annihilation cross section in Minkowski space-time, the infinite time limit leads to local Lorentz invariance. The threshold $\Theta$ function in (\ref{siginfi}) has the following origin: first we note that $\overline{X}(\eta)$ given by  eqn. (\ref{Xmin}) can also be written as
$\overline{X}(\eta) = (\mathcal{S}(\eta)-4m^2)/[( E_T(\eta) + K_{Tp}(\eta) )H(\eta)]$ where the denominator is always positive. Therefore, for $\mathcal{S}> 4 m^2$  as $H(\eta) \rightarrow 0$ it follows that $\overline{X}(\eta) \rightarrow \infty$  and the $X$  integral in eqn. (\ref{sigmasfin}) yields a non-vanishing result, whereas for $\mathcal{S}<4 m^2$ and  $H\rightarrow 0$ the lower limit $\overline{X}(\eta) \rightarrow -\infty$ and this integral vanishes in the ``long time limit''. Hence, the emergence of the sharp kinematic  threshold is a consequence of taking the infinite time limit. However,   the particle horizon $\eta-\eta_i$ is finite, consequently the cross section must be considered during a finite time interval, thereby allowing several  novel processes.

\vspace{1mm}

\subsubsection{Freeze-out of the cross section:} Consider the specific case of pair annihilation in the center of mass (CoM), namely with $K=0$ and with $M\ll m$, and in a time regime during which  the incoming particles feature a physical momentum $k_{ph}(\eta) > \sqrt{m^2-M^2}$ corresponding to the (CoM) total local energy  above the (physical) production threshold $2m$. As time evolves the scale factor grows and the physical wavevector is redshifted \emph{below} the threshold value for production of the heavier species, when this happens the total local energy falls below threshold and the integral in (\ref{sigmasfin}) begins to diminish, vanishing fast because $H(\eta)$ is becoming smaller with cosmological expansion.  We refer to this phenomenon as a \emph{freeze-out} of the production cross section.

\begin{figure}[!h]
\begin{center}
\includegraphics[height=4in,width=4.5in,keepaspectratio=true]{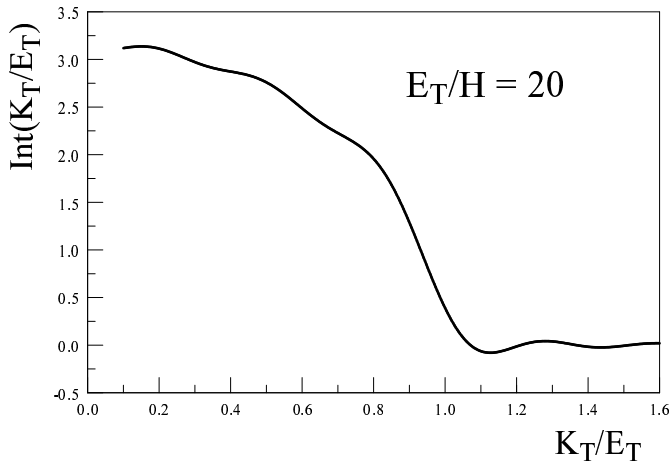}
\caption{  The function $Int(K_T/E_T)$ vs $K_T/E_T$ for $K_T=2m~;~E_T/H=20$.}
\label{fig:fgr1}
\end{center}
\end{figure}

 We emphasize that this freeze-out is different from the usual freeze-out of species as a consequence of the dilution of the particle density in a Boltzmann equation. Instead the freeze-out of the cross section is solely a consequence of considering the cross section during a  \emph{finite time }  including  explicitly the  time dependence of the kinematic threshold, local energy in terms of the redshifted momentum, and the Hubble radius. This phenomenon can be understood in the simpler case of $K=0$, (CoM)    from the integral in (\ref{sigmasfin}) which in this case can be written solely in terms of the ratios $K_T/E_T = 2m/E_T$ and $E_T/H$,

\be Int(K_T/E_T) = \int^{\infty}_{-\overline{X}(\eta)} \widetilde{\rho}(K_T/E_T;X\,H/E_T) \, \frac{\sin\big[X\big]}{X} \,dX   \,. \label{freeze} \ee    Figure (\ref{fig:fgr1}) displays $Int(K_T/E_T)$ vs. $K_T/E_T$ for $E_T/H =20; K=0$, for $K_T/E_T \ll  1$ the integral approaches $\pi \, \widetilde{\rho}(K_T/E_T;0)$ and vanishes for $K_T/E_T \gg 1$, confirming the above analysis.

Although fig. (\ref{fig:fgr1}) shows this phenomenon for a fixed ratio $E_T/H$, this ratio actually grows during cosmological expansion, the main conclusion is confirmed by this numerical example. As $H$ diminishes under cosmological expansion the contribution from $K_T/E_T > 1$ diminishes, and in the strict limit $H\rightarrow 0$ the step function determining the kinematic threshold in eqn. (\ref{inflimi}) emerges.

It is clear that taking the infinite time limit ($H=0$) too early will not capture this dynamics: if $k_{ph}$ is larger than the threshold value, the cross section grows rather than remaining constant as the step function in (\ref{inflimi}) would suggest, as the redshifted $k_{ph}$ falls below threshold it begins to diminish, vanishing rapidly but smoothly as $H\rightarrow 0$ rather than the abrupt vanishing suggested by the step function in (\ref{inflimi}).

\subsubsection{Anti-Zeno effect: production below threshold.}
The analysis above and the results displayed in fig.(\ref{fig:fgr1}) also suggest another  phenomenon consequence of the finite time, the possibility of pair production when the total (local) energy is below threshold, which is  evident in fig.(\ref{fig:fgr1}) in the tail  for $K_T/E_T > 1$.   To understand the origin of this phenomenon it is convenient to re-write the integral in (\ref{FGR}) in terms of the local energy and the Hubble rate, setting $\eta_i \rightarrow 0$,
\be \int^{\infty}_{-\infty} \rho(K_0;K) \, \frac{\sin\Big[\big(K_0 -\Omega_T(\eta)\big) \eta \Big]}{\big(K_0 -\Omega_T(\eta)\big)} \,dK_0 = \int^{\infty}_{-\infty} \rho\big[K_{0p}(\eta);K_p(\eta)\big] \, \mathrm{sinc}[K_{0p};E_T;H]
 \,dK_{0p}(\eta)\,, \label{intfgr}\ee where
 \be \mathrm{sinc}[K_{0p};E_T;H]=  \frac{\sin\Big[\big(K_{0p}(\eta) -E_T(\eta)\big)/ H(\eta) \Big]}{\big(K_{0p}(\eta) -E_T(\eta)\big)}\,,\label{sinc} \ee and  for two equal masses in the final state $m_3=m_4=m$
\be \rho\big[K_{0p}(\eta);K_p(\eta)\big] = \Bigg[1- \frac{4m^2}{K^2_{0p}(\eta)-K^2_p(\eta)} \Bigg]^{1/2} \,\Theta\big(K_{0p}(\eta)-K_{Tp}(\eta)\big) \, , \label{rhophys}\ee with
\be K_{0p}(\eta) = \frac{K_0}{a(\eta)}~~;~~ K_p(\eta)= \frac{K_p}{a(\eta)} ~~;~~ K_{Tp}(\eta) = \frac{K_T(\eta)}{a(\eta)}\,.\label{physpars}\ee

The \emph{sinc} function (\ref{sinc}) is strongly peaked at $K_0 =E$ with height $1/H$ and width $\simeq 2\pi H$.

\begin{figure}[!h]
\begin{center}
\includegraphics[height=4in,width=4.5in,keepaspectratio=true]{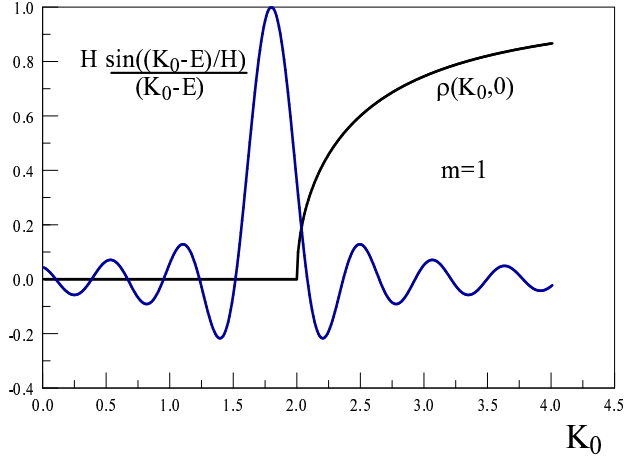}
\caption{    $\rho(K_0,0)$ and $H\,\sin[(K_0-E)/H]/(K_0-E)$ vs $K_0$ in units of $m$ ($m=1, K_T =2, K=0$) for $E_T \equiv E = 1.8$, $E/H =20$.}
\label{fig:antizeno}
\end{center}
\end{figure}

If $E_T$ is below threshold, but within a distance $\simeq \pi H$ from threshold, the ``wings'' of this function still overlap with the spectral density yielding a non-vanishing overlap integral. This is a manifestation of the phenomenon of threshold relaxation found within a different context in ref.\cite{herringdecay} and of the \emph{antizeno} effect which refers to the enhanced production as a consequence of uncertainty\cite{antizenokuri}. Such an effect has been   studied in quantum field theory in ref.\cite{boyaantizeno}.  Fig.(\ref{fig:antizeno}) shows both functions for $K=0$ in the case when $E_T$ is below the production threshold. The width of the oscillatory function is $\simeq 2\pi H$, the figure clearly shows that the ``wings'' of the \emph{sinc} (\ref{sinc}) function have a non-vanishing overlap with the spectral density.

The overlap between the  \emph{sinc}  function  (\ref{sinc}) and $\rho(K_0,K)$ is a consequence of \emph{uncertainty}, the Hubble scale $H$ introduces a  (physical) energy uncertainty associated with the time scale $1/H$. It is this energy uncertainty,   a consequence of the finite time scale $1/H$, which is reflected in the broadening of the   \emph{sinc} function that allows the overlap with the spectral density even when it is peaked below threshold thus allowing the production of more energetic states. This is the origin of the non-vanishing result for the integral (\ref{freeze}) for $K_T/E > 1$ displayed in fig. (\ref{fig:fgr1}).  This phenomenon has been recognized as the \emph{antizeno} effect in the quantum optics literature\cite{antizenokuri} and  has been observed in trapped cold sodium atoms\cite{antizenoobs} where the time uncertainty is introduced through the measurement process. In this case it is the inverse of   \emph{the particle horizon} ($\eta$ in comoving or $a(\eta)\eta =1/H =2\,t$ in physical coordinates), namely the age of the Universe,  which introduces the uncertainty. Under cosmological expansion the particle horizon increases, therefore the uncertainty  diminishes,  the  \emph{sinc} function becomes narrower and the overlap with the spectral density becomes smaller and eventually vanishes, therefore closing the uncertainty window for sub-threshold production.

The condition for the resulting integral to have a non-vanishing contribution and for significant below-threshold production is
\be K_T -E_T \simeq \pi \,H  \Rightarrow 4m^2 - \mathcal{S} = 2\pi\,E_T\,H \,,  \label{condibeth}\ee where we have considered $m_3=m_4=m$ in the final state and we kept the leading order in the adiabatic expansion for $E_T \gg H$.   In Minkowski space-time the threshold condition is $\mathcal{S}_M = 4 m^2$ therefore writing $\mathcal{S}=  \mathcal{S}_M - \delta \mathcal{S}$ the  finite particle horizon relaxes the threshold condition for production with
\be \delta \mathcal{S} = 2\,\pi E_T H \,. \label{relS} \ee It is noteworthy that the right hand side of  the condition (\ref{relS}) breaks local Lorentz invariance.

Since the integral in (\ref{sigmasfin}) vanishes   for $K_T - E_T \gg H$ and reaches the asymptotic infinite time limit for $E_T - K_T \gg H$ only a small window of width $\simeq H$ in energy and momentum yields sub-threshold production.  Writing the condition (\ref{condibeth}) as
\be K_T = E_T\Big(1+ \pi \frac{H}{E_T} \Big) \,,  \label{condibeth2}\ee
the long time  limit    $E_T/H \gg 1$ ($\Omega_T(\eta) \eta \gg 1$) invoked in the derivation of the cross section (\ref{sigmasfin}), implies that   $K_T \simeq E_T \gg H$ yielding the general condition for subthreshold production
\be   2m^2 -M^2 -k_{ph1}k_{ph2}-E_1E_2 \simeq \pi H\,E_T \,. \label{finicondi} \ee As a simple example, consider annihilation of the incoming particles in their  (CoM) with threshold $K_T = 2m$, subthreshold production of daughter particles would occur
with
\be k_{ph} = \Big[m^2-M^2-\pi\,m\,H\Big]^{1/2} \,,\label{kth}\ee for example with $m = 10^5\,\mathrm{GeV}, M = 10^2\,\mathrm{GeV}, H\simeq 10^2\,\mathrm{GeV}$ production of the heavier particle still occurs with  a value of $k_{ph}$ which is $\simeq 10^{2}\,\mathrm{GeV}$ below the corresponding threshold value in Minkowski space-time. This value of the Hubble rate during (RD) corresponds to an ambient temperature $T \simeq 10^{10}\,\mathrm{GeV}$ and $H/E_T \simeq 10^{-3}$.
Consider a more general example with $m \gg M,H$ and $k_{ph1} \gg k_{ph2} \gg M$, writing  $k_{ph2} = \alpha \, k_{ph1} $ with $\alpha \ll 1$, the inequality (\ref{finicondi}) is fulfilled for
\be k_{ph1} \simeq k_{M}-\frac{\pi H}{4\alpha} ~~;~~k_M = \Big[\frac{2m^2-M^2}{2\alpha}\Big]^{1/2} \,, \label{exa2} \ee  therefore for $\alpha \ll 1$ below threshold production occurs for values of $k_{ph1}$ which are smaller than that in Minkowski space-time ($k_M$) by $\pi H/4\alpha \gg H$. For the masses chosen above, and taking as an example $\alpha \simeq 10^{-2}$,  sub-threshold production consistent with $H/E \ll 1$ occurs for $H \lesssim 10^{3}\,\mathrm{GeV}  \Rightarrow a(\eta) \lesssim 10^{-24}$, again corresponding to an ambient temperature $T \lesssim 10^{10}\,\mathrm{GeV}$. An important consequence of this phenomenon is that if heavy dark matter is produced via pair annihilation of a much lighter species,   sub-threshold production implies an enhancement of the dark matter  abundance.

As cosmological expansion proceeds, $H$ diminishes and the window for sub-threshold production closes.  The important aspect of this analysis is that the finite value of $H$, namely a finite particle horizon $\propto 1/H$ provides an uncertainty allowing processes that would be forbidden by strict energy conservation.


In this case with only the ``wings'' of the $sinc$ function overlapping with the spectral density, it is possible (but not certain) that the transition rate becomes \emph{negative} during a brief transient period. Whether the transition rate becomes negative during a transient depends on momenta and the behavior of the spectral density near threshold. This important aspect is discussed in more detail in section (\ref{sec:discussion}). However, as in the previous section, we can integrate the transition rate within a finite  and short time interval during which the conformal energies do not vary much, to leading (zeroth) order in the adiabatic expansion we obtain
\be \int^{\eta}_0 \Gamma_{i\rightarrow f}(\eta')\,d\eta' = \frac{\lambda^2}{2V}\, \frac{1}{ \Omega_{k_1}(\eta)\Omega_{k_2}(\eta)}\,  \int^{\infty}_{-\infty} \rho\big[K_{0p}(\eta);K_p(\eta)\big] \, \Big[\mathrm{sinc}[K_{0p};E_T;H/2]\Big]^2
 \,dK_{0p}(\eta)\,. \label{intgami} \ee This result is manifestly positive in agreement with (\ref{Ptot}), and again similar to Fermi's Golden rule, however, now taking (formally) the limit $H \rightarrow \infty$ the bracket inside the integral in (\ref{intgami}) yields $ \pi\, H(\eta)\,\delta(K_{0p}-E_T(\eta))$ which would lead to a vanishing result since the spectral density vanishes for $K_{0p}=E_T(\eta)$ but multiplies $H \rightarrow \infty$. The total integral yields a finite and positive result, which however does \emph{not} grow secularly with conformal time.

\vspace{1mm}

\subsection{Scattering:}\label{subsec:scattering}

The results obtained above apply directly to the scattering process $\varphi \,\chi \rightarrow \varphi\,\chi$ with minor modifications, a combinatoric and symmetry factor for the final state resulting in $\lambda^2 \rightarrow \lambda^2/2$, different masses in the initial state $m,M$ and in  the flux factor,   and a different spectral density reflecting the two different masses of the particles in the initial and final states. Following the same steps leading up to (\ref{sigmasfin}) we now find
\be \sigma(\eta) = \frac{\lambda^2}{16\,\pi^2 \,a^2(\eta)\, \Big[\big(\mathcal{P}_1(\eta)\cdot \mathcal{P}_2(\eta)\big)^2-M^4 \Big]^{1/2} } \, \int^{\infty}_{-\overline{X}(\eta)} \widetilde{\rho}_{+}( X;\eta)\,\widetilde{\rho}_{-}( X;\eta) \,\, \frac{\sin\big[X\big]}{X} \,dX \,,\label{sigmascat} \ee where
\be \widetilde{\rho}_{\pm}( X;\eta)  =  \Bigg[1- \frac{(M\pm m)^2}{\mathcal{S}(\eta)+X^2 H^2(\eta)+2XE_T(\eta)H(\eta)}   \Bigg]^{\frac{1}{2}}~~;~~E_T(\eta) = (\Omega_{k_1}(\eta)+\omega_{k_2}(\eta))/a((\eta))\,,\label{rhopm}\ee and in this case
\be \overline{X}(\eta) =  \frac{ E_T(\eta) - K_{Tp}(\eta) }{H(\eta)}  ~~;~~ K_{Tp}(\eta)=\Big[K^2_{ph}(\eta)+ (M+m)^2 \Big]^{1/2} \,. \ee It is straightforward to confirm that, for collinear scattering
\be K_{Tp}(\eta) = \Big[ E^2_T(\eta) -2 (k_{ph1} k_{ph2}+ E_1 E_2-m M)\Big]^{1/2} \leq E_T \,.\label{noth} \ee
 We note that for $H\neq 0$ the cross section is \emph{not} invariant under the local Lorentz transformation,   again  as a consequence of the finite particle horizon, or equivalently finite time.

In the long time limit $H \rightarrow 0$ we find
\be \sigma_{ph}(\eta\rightarrow \infty) = \frac{\lambda^2\,\widetilde{\rho}_{+}( 0;\eta)\,\widetilde{\rho}_{-}( 0;\eta)}{32\,\pi \, \Big[\big(\mathcal{P}_1(\eta)\cdot \mathcal{P}_2(\eta)\big)^2-M^4 \Big]^{1/2} } \,  \,. \label{sigfisscat} \ee This is the result in Minkowski space-time, manifestly invariant under local Lorentz transformations  but with the kinematic variables depending adiabatically on time. We emphasize that this is an approximation that formally corresponds to taking the scale factor $a(\eta) \rightarrow \infty$, therefore leading  to an infinite redshift of the physical momenta that enter in the flux prefactor.

\section{Discussion}\label{sec:discussion}

\vspace{1mm}

\textbf{Subtleties of the transition rate:} In a finite time interval, the transition rate and cross section feature oscillations as a consequence of time-energy uncertainty. It is possible that within some range of (local) energy and momenta the transition rate may be negative in some circumstances. This is more explicit in the case of sub-threshold production analyzed in section (\ref{subsec:pair}). For this situation since the transition rate at large time vanishes, transient phenomena consistent with time-energy uncertainty yields the oscillatory behavior that probes the subthreshold region during a time interval and  may lead to negative values.  Whereas the total number of events (\ref{Ptot}) is manifestly positive, the event \emph{rate} may become negative during these transients.  This subtle behavior can be traced to the definition of the transition rate (\ref{rate}), which may indeed become negative during a finite time interval in some circumstances. This is not a feature of the cosmological setting  as it is also the case in Minkowski space-time analyzed in appendix (\ref{app:Xminkfiti}) which coincides  in the infinite time limit   with the usual definition as the total transition probability, (which grows linearly in time at long time), divided by the total time elapsed. There are at least two arguments in favor of the definition (\ref{rate}): \textbf{a:)} the total number of detected events is generally obtained as
\be N = \sigma \,L ~~;~~ L = \int \mathcal{L}(t) dt \,,\label{toteven}\ee where $L$ is the integrated luminosity and the luminosity $\mathcal{L}(t)$ has units of $cm^{-2}\,{s}^{-1}$. Obviously such definition of the number of events at a detector assumes a time independent cross section, and from this definition $\sigma = N/L$ is manifestly positive. Allowing the cross section to depend on time, the total number of events should, consequently, be defined as
\be N = \int \mathcal{L}(t)\,\sigma(t) dt  \Rightarrow \frac{dN}{dt} = \sigma(t) \mathcal{L}(t)\,. \label{totevent} \ee   This definition is consistent with eqn. (\ref{rate}). \textbf{b:)} consider a process $1+2 \rightarrow 3+4$, the gain term in the Boltzmann equation for the distribution function of particle $3$ is
\be \frac{d f_3(t)}{dt} = \int d[1]\int d[2] f_1(t)\,f_2(t) (\sigma(t) v_{12}) \label{boltz3} \ee with $v_{12}$ the relative velocity. Therefore the definition of $\sigma(t)$ in terms of the transition rate given  eqn. (\ref{rate}) is consistent with the usual Boltzmann equation.


Although it remains to be shown that  the transition rate in the appropriate Boltzmann equation in cosmology is determined by $\sigma \,v_{12}$ (however  this is the usual
formulation), the main point of eqn. (\ref{boltz3}) is that it explicitly implies that the \emph{rate} of change of the population  is defined as its \emph{time derivative}  namely $df/dt$, consistently with our definition for the transition rate (\ref{Ptot}).

One may propose an alternative definition of the transition rate which is manifestly positive as the time average (now in conformal time) (taking $\eta_i\rightarrow 0$)
\be \overline{\Gamma(\eta)}  \equiv \frac{1}{\eta} \int^{\eta}_{0} \Gamma(\eta') \,d\eta' \,,\label{avegam}\ee where the integral is given by (\ref{intgami}),  and define a time-averaged cross section by dividing by an instantaneous unit flux at time $\eta$  in a long time limit. In cosmology such a definition would not be consistent: the integral in (\ref{avegam}) depends on the expansion history, and multiplying by the flux at a late time when the physical momenta and relative velocity had undergone a large redshift would not describe the cosmological evolution consistently.  Therefore, we conclude that despite the subtleties and the counterintuitive possibility of a negative transition \emph{rate } mainly in the case of subthreshold phenomena,  the definition of the rate (\ref{rate}) is consistent with unitarity, the optical theorem, a manifestly positive \emph{total} number of events, the usual relation between the total number of events at a detector in terms of the integrated luminosity, and the Boltzmann equation.  The time integral of the transition rate (\ref{rate}) yields the total transition probability that is manifestly positive.  The cross section \emph{per se} is an ingredient whose  time integral in combination with flux factors or distribution functions yields the total number of events  which is positive (eqn. (\ref{Ptot})).   Furthermore, the phenomena discussed above such as local Lorentz violation and sub-threshold production yielding contributions  to events that would be forbidden by strict energy conservation will remain features of the time integrated quantities.

\vspace{1mm}

\textbf{General lessons and possible cosmological consequences:}

\begin{itemize}

\item \textbf{Threshold relaxation and energy uncertainty:}
Although our study above has focused on a model local interaction, there are many general lessons that we can draw from it. To begin with, the lack of strict energy conservation quantified by the uncertainty $H$ in local physical energy variable is a direct result of the finite particle horizon or, equivalently, finite time. Comparison with the analysis in  Minkowski space-time but during a finite time interval highlights that this feature is indeed quite generic. The S-matrix approach is always formulated with in-states prepared in the infinite past and out states measured in the infinite future, clearly warranted in typical experimental situations. However, is unwarranted during the early stages of cosmological history with a finite particle horizon. A direct consequence of the (local physical) energy uncertainty $\simeq H$ is that kinematic thresholds that reflect strict energy momentum conservation are relaxed within a window of order $H$ thereby allowing processes that would be otherwise forbidden by strict energy conservation. This is the case for sub-threshold production studied above. This is a fundamental and generic feature of processes in cosmology and are independent of the particular interaction between the various fields.

As an example of potential cosmological importance of this phenomenon, let us consider that post-inflation reheating occurs at a high energy/temperature scale, for example intermediate between the GUT and the Standard Model scales, $T \simeq 10^9\,\mathrm{GeV}$. This temperature corresponds to a scale factor $a(t) \simeq 10^{-20}$ and a  Hubble expansion rate $H \simeq \mathrm{GeV}$, which determines the energy uncertainty that characterizes threshold relaxation. This uncertainty window would allow scattering or production processes that would be forbidden by strict energy conservation to produce degrees of freedom near the scale of this uncertainty. Reheating at a higher energy/temperature scale would widen this uncertainty window.


Consequences of energy uncertainty within a finite time interval have been recently discussed also within different contexts\cite{recai,arza}.

\vspace{1mm}

\item \textbf{Violations of local Lorentz invariance:} We recognized a violation of Lorentz invariance as a  consequence of the finite time analysis, both with cosmological expansion as well as Minkowski space-time  during a finite time interval (see appendix (\ref{app:Xminkfiti}))  . In Minkowski space-time such violation of Lorentz invariance is expected, a finite time interval is not a Lorentz invariant concept. Within the context of an expanding cosmology, a finite time analysis of interaction rates and cross sections is a necessity: an infinite time limit as implied in the S-matrix formulation, ignores both formally and conceptually   the different stages in the expansion history,     the cosmological redshift of physical wavevectors and the finiteness of the particle horizon. It is this necessity to consider processes directly in real and finite time that leads unequivocally to violation of local Lorentz invariance. At heart, a finite time analysis  is simply a recognition of a finite particle horizon $\propto 1/H$ in physical coordinates, which introduces an energy uncertainty $\propto H$. In appendix (\ref{app:Xminkfiti}) we obtained the cross section for pair annihilation during a finite time interval in Minkowski space-time, which also displays a similar violation of Lorentz invariance, see eqns. (\ref{sigrhot},\ref{rhothe}).  However, in a ``terrestrial'' experiment the energy and time scales justify taking the infinite time limit: with typical time scales of $10^{-4}\,\mathrm{secs}$ (a beam travelling few $ \mathrm{km}$ until detection) the typical energy uncertainty is $\simeq 10^{-10}\,\mathrm{eV}$ and typical detector momentum resolutions  $\simeq \mathrm{KeV}-\mathrm{MeV}$ the finite time corrections are  experimentally  completely  negligible, justifying the infinite time limit with the concomitant Lorentz invariant cross section.

 Possible sources of Lorentz violations have been advocated within the context of quantum gravity and Planck scale physics\cite{LV1,LV2,LV3,LVkoste} with possible consequences for CP and CPT violation. Our study here shows that Lorentz violations emerge naturally as a consequence of considering fundamental processes in an expanding cosmology directly in real time and accounting for the finite particle horizon. The example of pair annihilation displays the consequences of the finite time in the form of the freeze-out and antizeno effects that lead to small but non-vanishing sub-threshold production of heavy particles as analyzed in the previous section. The freeze-out of a production cross section as a consequence of the redshift of wavevectors below the threshold, is, similarly, an inescapable consequence of the cosmological expansion, hence it is a generic feature associated with the finite time evolution. These phenomena are directly correlated with the violation of local Lorentz invariance. To be sure, within the adiabatic approximation these effects are small and the window for their impact closes with the cosmological expansion. However, during the time that this window remains open, the Lorentz violating processes \emph{may} be of importance for  {CP} violation and perhaps baryogenesis. The possible impact of these Lorentz violating aspects deserve further scrutiny.

\vspace{1mm}

\item \textbf{Impact on quantum kinetics:} An important cosmological application of the transition rate (\ref{rate}) and cross section,  is to input these into a quantum kinetic Boltzmann equation for the distribution function of particles. A kinetic equation for the distribution function of a $\varphi$ particle $\mathcal{F}_{\varphi}(\vk,\eta)$\ for example, is of the form $d\,\mathcal{F}_{\varphi}(\vk_{1},\eta)/d\eta = \mathrm{gain}-\mathrm{loss}$  where the ``loss'' term inputs the transition rate (\ref{rate}) and the gain term inputs the rate for the reverse process $\chi \chi \rightarrow \overline{\varphi}\varphi$. An important feature of quantum kinetic equations is detailed balance which leads to  local thermodynamic equilibrium as a fixed point of the kinetic equation. However, detailed balance depends crucially on energy-momentum conservation both in the gain and loss terms\cite{kolb,bernstein,dodelson}. The results of the previous section indicate that there \emph{may} be possible important modifications arising from the energy uncertainty $\propto H$ and lack of energy conservation. These could lead to modifications of detailed balance and consequently introduce novel non-equilibrium effects, which along with the violation of local Lorentz invariance could, potentially,  be important for baryogenesis and leptogenesis, even when such effects are
    small, of  $\mathcal{O}(H)$ as compared to temperature or energy scales. These intriguing possibilities deserve  further and deeper study.

\vspace{1mm}

\item \textbf{Resonant cross sections smoothed by a finite particle horizon:} The energy uncertainty introduced by $H$ brings an interesting possibility for the case of resonant cross sections, where the resonance is a consequence of a massive particle exchange as an intermediate state. In S-matrix theory in a resonant reaction the intermediate state goes on-shell and the enhancement of the cross section is a consequence of this intermediate state propagating over long time scales, a pole in the cross section is actually smoothed out
by the lifetime of the intermediate state that has gone ``on-shell''. The (Breit-Wigner) width of the cross section is a manifestation of the lifetime of the intermediate state. We conjecture that a similar case of an intermediate stage ``going on its mass shell''  in an expanding cosmology only propagates during the particle horizon. Therefore for a very long-lived intermediate state during the regime when the particle horizon is smaller than the lifetime,  we expect a resonant cross section will feature a width $\propto H$ rather than the natural width of the decaying state. As $H$ diminishes, the width of the intermediate state will replace $H$ in the broadening of the cross section. This possibility would have potentially interesting consequences in dark sectors that are connected to the Standard Model via mediators, when such mediators could lead to resonant cross sections. We are currently studying this scenario.

\item \textbf{No corrections to Big Bang Nucleosynthesis:} Although our study has focused on a quartic contact interaction among bosonic fields and does not apply directly to the case of (BBN), one of the general lessons drawn from this study is the dependence on the various energy and time scales. The effects of finite time such as local  Lorentz violation and threshold relaxation in production cross sections  are phenomena associated with the ratio $(K_T -E)/H$. (BBN) occurs during the (RD) era at a time scale $t\simeq \mathrm{few~~minutes}$ and temperatures $T \simeq \mathrm{few~~MeV}$ corresponding to $H \simeq 10^{-24}\,\mathrm{GeV}$ with nucleon energies $\simeq \mathrm{GeV}$ and binding energies $\simeq \mathrm{few ~~MeV}$ therefore the relevant ratio is $E/H \simeq 10^{27}$ which leads to   the energy conserving limit of the cross section. Therefore, without a doubt the interesting phenomena associated with finite time do not modify the standard results of (BBN).

    \item \textbf{When is the S-matrix approximately reliable?, and when is it not?} The study of the previous sections along with the main results invites the above questions.  The adiabatic approximation described in this study and the comparison with the results of a finite time analysis in Minkowski space-time, all suggest that, as expected, for a Hubble rate $H(t)$ much smaller than the typical local energies, the uncertainty $\propto 1/H$ can be safely neglected. Taking the infinite time limit in the transition amplitudes, \emph{but} allowing the redshift of momenta in the local energies of the external particles and in reaction thresholds, which lead to the freeze-out of the cross sections,  yields a reasonably reliable approximation. Such is the case for (BBN). However, there are at least two scenarios wherein the S-matrix approach will ultimately be unreliable: very early during (RD) when $H$ is large, comparable with the mass scales of particles, or whenever the adiabatic approximation breaks down. The first instance would be relevant for a description of thermalization during reheating if it occurs at a high energy scale. The second scenario applies in the case  of super-Hubble wavelengths for particles with masses smaller than the Hubble rate. Under these circumstances, quantization with the full mode functions such as the parabolic Weber functions in the (RD) case is required, and even the program described in this study which relies on the validity of the adiabatic approximation would need a reassessment.

    \end{itemize}

\vspace{1mm}

\textbf{On wave packets:}

 As discussed in elementary textbooks,  a (more)  correct description of scattering and or production should invoke a treatment of initial states in terms of wave packets localized in space-time. While formally correct, practically all calculations describe initial states as Fock eigenstates of momentum, namely plane waves, disregarding the fact that these are not localized. Invoking wave packets, while addressing the issue of locality of the initial state presents several conceptual and technical challenges even in Minkowski space-time: wave packets spread through dispersion since each wave vector evolves differently in time, this, by itself prevents a formal infinite time limit. The localization in space and in momentum components must be assumed to be such that the incoming ``beams'' are localized in space-time but yet be nearly monochromatic so the momenta of the initial and final states which are the labels of the S-matrix elements, are fairly well defined. Last, but by no means least, wave packets do not transform as irreducible representations of the Lorentz group, thereby breaking Lorentz invariance. These are well known shortcomings of a wave packet description. A thorough analysis of many of these issues within the context of the asymptotic formulation of the S-matrix has been discussed recently in ref.\cite{collins}.  In a cosmological space-time these conceptual and technical subtleties of the wave packet formulation are compounded by two important aspects: i) each Fock momentum state in the wave packet components evolves with a non-local phase in time, ii) the cosmological redshift of the momenta. The width in momentum (or localization length in space) also determines an uncertainty. All these aspects introduce yet several more layers of technical complications which
 ultimately need to be addressed. These notwithstanding, the general lessons and  novel phenomena described above are overarching and robust and qualitatively not affected by a wave packet treatment.

\section{Conclusions and further questions}\label{sec:conclusions}
Motivated by the importance of transition rates and cross sections in a wide range of processes in the early Universe, this article is devoted to obtaining these fundamental ingredients from first principles in a spatially flat, radiation dominated cosmology. The main objectives are to  re-assess  the usual S-matrix approach, which takes the infinite time limit, as applied to the cosmological setting, to highlight its shortcomings and provide a systematic framework that includes consistently the cosmological expansion, the finite particle horizon and  to  explore their consequences. We begin this program by focusing on bosonic fields interacting via a quartic local contact interaction, a simpler setting which, however, allows us to study various important processes  such as  production of a heavier species and  scattering, and yields important  and more general lessons. Field quantization in the cosmological curved space-time explicitly shows the conceptual and daunting technical obstacles to implementing the usual approach to  transition rates and cross sections. To overcome these obstacles we introduce a physically motivated adiabatic approximation  that relies on a separation of time scales encoded in a small ratio $H(t)/E(t)$ with $H(t)$ the Hubble expansion rate and $E(t)$ a typical particle energy measured by a local observer. The smallness of this ratio corresponds to typical wavelengths (either de Broglie or Compton) much smaller than the particle horizon at a given time and also to a wide separation between the (shorter) microscopic time scale of oscillations $\propto 1/E$ and the (longer) time scale of expansion $\propto 1/H$. We show that the leading (zeroth) adiabatic order dominates the transition rates and  obtain the cross section for various processes during  a finite time interval consistently with a finite particle horizon. We compare these cross sections to those in Minkowski space-time within a finite time interval. We find several novel effects associated with the finite particle horizon in the case of production of a heavier species from the annihilation of a lighter species: \textbf{i)} Freeze-in of the production cross section:  the cosmological redshift of physical momenta eventually diminishes it below the production threshold, keeping a finite time in the cross section reveals that it vanishes fast but continuously when this happens and production shuts off. \textbf{ii)} A finite particle horizon ($1/H$ in physical coordinates) determines an uncertainty in the (local) energy of order $H$ allowing processes that would be forbidden by strict energy conservation. In particular this uncertainty allows \emph{sub-threshold} production of heavier particles within a window of width $\simeq H$ in (local) energy, during a small interval of time. If a heavy dark matter particle is produced via the pair annihilation of a much lighter species, sub-threshold production may lead to a larger dark matter abundance after the cross section freezes-out.

We also find the important result that allowing for the finite particle horizon leads to a \emph{violation of local Lorentz invariance}, this is a general result for all processes and is a direct consequence of explicitly keeping the cosmological time evolution in the transition rates.

An important corollary of our study is the limitation of the adiabatic approximation: processes that involve wavelengths that are larger than the particle horizon and   masses $m \ll H$ at a given time must be studied non-perturbatively with the full mode functions, for bosonic particles these are linear combinations of Weber's parabolic cylinder functions given by (\ref{fconf}).

We comment on possible implications of these results, for example their impact on CP and CPT violation, quantum kinetics,  baryogenesis and/or leptogenesis as well as possible new phenomena associated with resonant cross sections. The adiabatic approximation advocated in this article should pave the way towards a deeper understanding of   interactions of the Standard Model in the regime of validity of such approximation.

Caveats with the definition of a transition rate during finite time intervals have been identified and discussed. We explored alternative definitions, which, however are inconsistent with the time dependence of energy and cosmological redshift of physical momenta. These aspects should motivate a further exploration of the concept of transition rates including transient phenomena in early Universe cosmology.

Furthermore, we have discussed under what circumstances a modified  S-matrix approach is approximately reliable, and when it is not, and in the latter case we have identified the proper quantization procedure.

In summary, this study has revealed hitherto unexplored consequences of considering these fundamental processes in the early Universe with proper account of the finite particle horizon i.e, finite time,  in their description. The results open new avenues of inquiry that will be the subject of forthcoming studies.

\appendix

\section{Cross section in Minkowski space-time in a finite time interval: }\label{app:Xminkfiti}
The S-matrix calculation of a cross section begins with the transition amplitude from $t= -\infty$ to
$t=\infty$. In order to compare to the analysis of cross section during a rapidly evolving cosmology to the Minkowski space-time case, in this appendix we obtain the cross section in Minkowski space-time but during a finite time interval.

\subsection{Pair annihilation:}

 The flat space-time case is obtained by replacing $\eta \rightarrow t$,   setting $a(\eta) \equiv 1$, and  the frequencies $\Omega_k,\omega_p$ are time independent. The discussion leading up to the definition of the interaction rate, eqn. (\ref{rate}) is exactly the same but with the above replacements. The transition matrix element  is now given by

 \bea  \langle f|H_I(t)|i\rangle  =    \frac{-i\,\lambda}{2\,V}\,\delta_{\vk_1+\vk_2 - \vp -\vq}~~\int^{t}_{t_i}   ~~\frac{e^{-i  \big(\Omega_{k_1} +\Omega_{k_2} -\omega_{p} -\omega_{q} \big) (t-t_1)}
}{\Big[ \Omega_{k_1}\, \Omega_{k_2}\, \omega_{p}\, \omega_{q}\Big]^{1/2}} ~  dt_1\,.  \label{anniampmink} \eea
 Following the steps leading up to eqn. (\ref{Ptot}) we find the total transition probability
  \be \mathcal{P}_{tot}(t) =   \int^t_{t_i} dt_2 \int^{t_2}_{t_i} dt_1 \,2\, \mathrm{Re}\Sigma[t_2;t_1]\,, \label{proptotmink}\ee where
  \be \Sigma[t_2;t_1] = \frac{\lambda^2}{8V\Omega_{k_1}\Omega_{k_2}} \int \frac{d^3p}{(2\pi)^3}\, \frac{{e^{i \big(\Omega_{k_1} +\Omega_{k_2} -\omega_{p} -\omega_{q}  \big) (t_2-t_1)}}}{\omega_p\,\omega_q}  ~~;~~ q = |\vec{k}_1+\vec{k}_2-\vec{p}| \,. \label{sigmink}\ee

 Therefore the transition rate (\ref{rate}) is given by
\be \Gamma_{i\rightarrow f}(t) = \frac{d\,\mathcal{P}_{tot}(t)}{dt} = 2 \int^{t}_{t_i} dt_1 \mathrm{Re}\Sigma[t;t_1]\,, \label{rate2mink}\ee and the time dependent cross section for one incoming particle is
\be \sigma(t)= \frac{\Gamma_{i\rightarrow f}(t)}{\Big(\frac{v_{rel}}{V}\Big)}\,. \label{minkX}\ee We now consider massless $\chi$-particles, in order to compare to the cosmological result. We will obtain the cross section in two different manners: i) integrating first in momentum, ii) integrating first in time.

\vspace{1mm}

\textbf{i:)} Integrating first in momentum, we need the integral (with a convergence factor $\epsilon \rightarrow 0^+$)
\be I[t-t_1] = \int \frac{d^3p}{(2\pi)^3}\, \frac{{e^{i \big(\Omega_{k_1} +\Omega_{k_2} -p -|\vec{k}-\vp| \big) (t_2-t_1)}}}{p\,|\vec{K}-\vp|}  = \frac{-i}{4\,\pi^2} \frac{e^{-i K(t-t_1)}}{t-t_1-i\epsilon} ~~;~~ \vec{K} = \vk_1+\vk_2 \,. \label{mom1st} \ee Gathering terms we find
 cross section during a finite time interval in flat space-time
\be \sigma(t) = \frac{\lambda^2}{16 \,\pi \, \Omega_{k_1}  \, \Omega_{k_2}  v_{rel}  } \,\,  \frac{1}{2}\Bigg[ 1 + \frac{2}{\pi} \int^{t}_{t_i}\, \frac{\sin[(\Omega_T -K)(t-t_1)]}{t-t_1} \,  dt_1 \Bigg]\,, \label{XsecMink}\ee where
\be   \Omega_T =  \Omega_{k_1}+\Omega_{k_2} ~~;~~ K = |\vec{K}|=|\vk_1+\vk_2|\,, \label{Jmink}\ee and for collinear scattering
\be \Omega_{k_1}  \,\Omega_{k_2}  v_{rel} = \Big[(P_1 \cdot P_2)^2- M^4\Big]^{1/2} \label{flumink}\ee is the usual flux factor for two body scattering, with the four momenta $P^\mu_{1,2} = ( \Omega_{k_{1,2}}, \vk_{1,2} )$. The time integral in eqn. (\ref{XsecMink}) yields the sine integral function, therefore  setting $t_i\rightarrow 0$ we find
\be \sigma(t) = \frac{\lambda^2}{16\pi\,\Big[(P_1 \cdot P_2)^2- M^4\Big]^{1/2}}\,\,\frac{1}{2}\Big[1+ \frac{2}{\pi}Si[( \Omega_T -K)\,t] \Big] \,.  \label{sigminkfina}\ee In the infinite time limit $t  \rightarrow \infty$ with $Si[\infty] = \pi/2$ we find the correct Lorentz invariant annihilation cross section $\varphi^+ \varphi^- \rightarrow \chi \chi$ for massless $\chi$ as obtained from S-matrix theory. We note, however, that in a finite time interval the cross section is \emph{not} Lorentz invariant, this is of course expected as a finite time interval is not a Lorentz invariant concept.

\vspace{1mm}

\textbf{ii:)} Integrating first in time, with $\Omega_T = \Omega_{k_1}+\Omega_{k_2}~;~K = |\vk_1+\vk_2|$ and the result (\ref{flumink}) for the flux factor for collinear scattering, we find
\be \sigma(t)= \frac{\lambda^2}{ 4\,\Big[(P_1 \cdot P_2)^2- M^4\Big]^{1/2} }\, \int \frac{d^3p}{(2\pi)^3}\frac{\sin\Big[\Big(\omega_p+\omega_q - \Omega_T \Big)t\Big]}{\omega_p \,\omega_q \, \Big(\omega_p+\omega_q - \Omega_T \Big)} \,, \label{sigtimefirst} \ee which we write  as

\be \sigma(t)= \frac{\lambda^2}{4 \Big[(P_1 \cdot P_2)^2- M^4\Big]^{1/2} }\, \int^{\infty}_{-\infty} \rho(K_0,K) \frac{\sin\Big[\Big(K_0- \Omega_T \Big)t\Big]}{   \Big(K_0 - \Omega_T \Big)} \,dK_0 \,, \label{sigrho} \ee where

\be \rho(K_0,K) = \int \frac{d^3p}{(2\pi)^3}\frac{\delta \big[ \omega_p+\omega_q - K_0  \big]}{\omega_p \,\omega_q }\,,  \label{rho} \ee is the two body Lorentz invariant phase space. For massless particles it is given by
\be \rho(K_0,K) = \frac{1}{4\pi^2} \Theta(K^2_0 -K^2)\,\Theta(K_0)\,.  \label{lips} \ee Introducing this result into (\ref{sigrho}) we find
\be \int^{\infty}_{-\infty} \rho(K_0,K) \frac{\sin\Big[\Big(K_0- \Omega_T \Big)t\Big]}{   \Big(K_0 - \Omega_T \Big)} \,dK_0 = \int^{\infty}_{-(\Omega_T-K)t} \frac{\sin(x)}{x} \, dx = \frac{\pi}{2}\Bigg[1+ \frac{2}{\pi}\, Si[(\Omega_T-K)t] \Bigg] \,,\label{sinint}\ee finally yielding
\be \sigma(t)= \frac{\lambda^2}{16\pi\, \Big[(P_1 \cdot P_2)^2- M^4\Big]^{1/2} }~~ \frac{1}{2}\Bigg[1+ \frac{2}{\pi}\, Si[(\Omega_T-K)\,t] \Bigg]\,, \label{sigt2}\ee which agrees with (\ref{sigminkfina}).

This result can be generalized to the case in which the final state corresponds to two particles of different masses $m_3,m_4$ respectively. In this case
 \be \rho(K_0,K) = \int \frac{d^3p}{(2\pi)^3} \frac{\delta\big(k_0 -\omega_3(p)-\omega_4(q)\big)}{\omega_3(p)\,\omega_4(q)} ~~;~~ q = |\vec{p}-\vec{k}| \,,  \label{rhomass}\ee is the Lorentz invariant two body phase space, it is given by
\be \rho(K_0,K) = \frac{1}{4\pi^2} \Bigg[1- \frac{(m_3-m_4)^2}{K^2_0-K^2} \Bigg]^{1/2}\,\Bigg[1- \frac{(m_3+m_4)^2}{K^2_0-K^2} \Bigg]^{1/2} \,\Theta\big(K_0-K_T\big) \, , \label{2prho}\ee where the
threshold $K_T$ is
\be K_T = \sqrt{k^2+(m_3+m_4)^2} \,.  \label{threshold} \ee
For the case of pair annihilation $\overline{\phi}\phi \rightarrow \chi \chi$ with $m_3=m_4 =m$, introducing the result (\ref{2prho}) into the expression (\ref{sigrho}) and defining $X= (K_0-\Omega_T)t$ we find
\be  \sigma(t)= \frac{\lambda^2}{16\,\pi^2 \,\Big[(P_1 \cdot P_2)^2- M^4\Big]^{1/2} }\, \int^{\infty}_{-\overline{X}(t)} \widetilde{\rho} (\mathcal{S},X;t)  \frac{\sin\big[X\Big]}{X} \,dX \,, \label{sigrhot} \ee where
\be \widetilde{\rho}_{\pm}(\mathcal{S},X;t)  \Bigg[1- \frac{4m^2}{\mathcal{S}+\frac{X^2}{t^2}+2\Omega_T\frac{ X}{t}} \Bigg]^{1/2} ~~;~~\overline{X}(t) = (\Omega_T-K_T)t   \,,  \label{rhothe}\ee and $  \mathcal{S} = \Omega^2_T - K^2$ is the usual Mandelstam Lorentz invariant. Note that for finite time the cross section is not Lorentz invariant, and it becomes Lorentz invariant in the $t \rightarrow \infty$ limit.

The integrated transition rate
\be \int^t_0 \Gamma_{i\rightarrow f}(t')\,dt'  =   \frac{\lambda^2}{2 \Omega_{k_1}\,\Omega_{k_2} \,V }\, \int^{\infty}_{-\infty} \rho(K_0,K) \Bigg[\frac{\sin\Big[\Big(K_0- \Omega_T \Big)t/2\Big]}{   \Big(K_0 - \Omega_T \Big)}\Bigg]^2 \,dK_0 \,, \label{interatemink} \ee which is recognized as the Fermi's golden rule result for the total transition probability as a function of time a manifestly positive result in agreement with equation (\ref{Ptot}).

\section{Time dependent kernels from spectral densities: }\label{app:specdens}

Let us consider the time dependent kernel
\be I(T;K) = \int \frac{d^3p}{(2\pi)^3} \frac{e^{-i\big(\omega_3(p)+\omega_4(q)\big)T}}{ \omega_3(p)\, \omega_4(q)} ~~ \omega_3(p) = \sqrt{p^2+m^2_3}~~;~~\omega_4(q)=\sqrt{|\vec{p}-\vec{K}|^2+m^2_4}\,, \label{kernel} \ee which can
be written in terms of the spectral density $\rho(K_0;K)$ given by eqn. (\ref{rhomass}) as
\be I(T;k) = \int^{\infty}_{-\infty} \,\rho(K_0;K)\,e^{-iK_0 T} dK_0 \label{specrepmink}\ee where the spectral density is given by eqns. (\ref{rhomass},\ref{2prho}).  The $K_0$ integral in (\ref{kernel}) can be carried out explicitly in few cases by introducing a convergence factor $T\rightarrow T-i\epsilon$ with $\epsilon \rightarrow 0^+$.

\vspace{1mm}

\textbf{i:) $m_3 =m_4=0; K_T =K $}
\be I(T;k) = -\frac{i}{4\pi^2}\,\frac{e^{-iK\,T}}{( T-i\epsilon )} = -\frac{i}{4\pi^2}\, {e^{-iK\,T}}\,\Bigg[ \mathcal{P}\Big(\frac{1}{T}\Big) + i \pi \delta(T) \Bigg]\,. \label{masslesskernelapp}\ee

\vspace{1mm}

\textbf{ii:) $m_3 =0, K=0;   m_4 \equiv m = K_T$ }

\be I(T;0) = -\frac{i}{4\pi^2}\,\frac{e^{-im\,T}}{( T-i\epsilon )}- \frac{m^2\,T}{4\pi^2} \Bigg[\frac{e^{-im\,T}}{m\,T}+i \Big[Ci[m\,T]+i\,\frac{\pi}{2}-i\,Si[m\,T] \Big] \Bigg] \,,\label{onemasszero}\ee where $Ci,Si$ are the cosine and sine integral functions respectively. The asymptotic behavior for short and long time are given by
\bea I(T \rightarrow 0;0)  & = &  -\frac{i}{4\pi^2}\,\frac{e^{-im\,T}}{( T-i\epsilon )} + \mathrm{constant} \,, \label{shorti} \\
I(T \rightarrow \infty;0)  & = &  -\frac{1}{8\pi^2}\,\frac{e^{-im\,T}}{m\,T^2}\,.  \label{longti1massive} \eea

\vspace{1mm}

\textbf{General Results:} Although the explicit form of the kernel may only be found in special cases,
the short and long time limit can be extracted generally in all cases.

\vspace{1mm}

\textbf{Short time limit:} the short time limit $T \rightarrow 0$ receives contributions from the full integral in $K_0$, however $\rho(K_0;K) \rightarrow 1/4\pi^2$ as
$K_0 \rightarrow \infty$, hence the integral is dominated by the large $K_0$ region which is ultraviolet divergent. This divergence is manifest as a singularity $\propto 1/T $ as $T\rightarrow 0$, which can be extracted by subtracting the spectral density by writing
\be \rho(K_0,K) = \frac{1}{4\pi^2}\Theta(K_0-K_T) + \widetilde{\rho}(K_0,K) ~~;~~ \widetilde{\rho}(K_0,K) = \rho(K_0,K)-\frac{1}{4\pi^2}\Theta(K_0-K_T)\,. \label{rhotil} \ee Since $\widetilde{\rho} \rightarrow 1/K^2_0$ as $K_0 \rightarrow \infty$ its integral yields a finite contribution to the time kernel as $T \rightarrow 0$. The integral of the first term yields a contribution similar to that of the massless case (\ref{masslesskernelapp}) but with $K \rightarrow K_T$. The integration of the second term, with $\widetilde{\rho}$ yields a finite contribution as $T\rightarrow 0$ since the integrands falls off as $1/K^2_0$ for large $k_0$ thereby ensuring the convergence of this integral in the short time limit. Therefore, we find the short time limit
\be I(T,K) = -\frac{i}{4\pi^2}\,\frac{e^{-i\,K_T\,T}}{( T-i\epsilon )} + \mathrm{finite} ~~;~~ T \ll m_{3,4} \,. \label{shortigen} \ee

\vspace{1mm}

\textbf{Long time limit:} the long time limit can be reliably extracted as follows. If the spectral density is \emph{analytic} near threshold, namely all derivatives are finite there, an asymptotic expansion in
\emph{integer inverse powers} of $T$ can be obtained. In the integral representation with the spectral density (\ref{specrep}) write
\be e^{-iK_0\,T} = \frac{i}{T}\,\frac{d}{dK_0} \Big(e^{-iK_0\,T} \Big)\,, \label{trick}\ee and integrate by parts, upending a convergence factor $T \rightarrow T-i\epsilon~;~\epsilon \rightarrow 0^+$ the upper limit of integration does not contribute and only the threshold does. If the spectral density vanishes at threshold as $\rho(K_0,K) \simeq [K_0-K_T]^{n}$ with $n$ an integer, the procedure (\ref{trick}) can be iterated yielding  the leading order
behavior at long time $K_T\, T \gg 1$  is
\be I(T;K) \propto \frac{1}{K^n_T\, T^{n+1}} ~~;~~ K_T \, T \gg 1 \,. \label{longtiint}\ee

This is the case for both massless particles in the out state, with $n=0$ and for only one massless particle and one massive in the out state, with $n=1$, which yield the results (\ref{masslesskernel}) and (\ref{longti1massive}) respectively. In the first case $\rho(k_0,k)$ is constant at threshold and in the second case $\rho(K_0,K) = (K_0-m)(K_0+m)/(4\pi^2\,K^2_0)$ and vanishes linearly at threshold $K_T =m$. The case when derivatives of $\rho(K_0,K)$ at threshold are singular requires a slightly different treatment.

Let us illustrate it  by considering the case $m_3=m_4 = m; k=0$    for which $\rho(k_0,0)$ features a square root singularity at threshold,
\be \rho(K_0,0) = \frac{1}{4\pi^2}\, \Bigg[1- \frac{4m^2}{K^2_0} \Bigg]^{1/2}\,.  \label{rho2masseq}\ee
 Introducing the change of variables
\be k_0 = 2m+ \frac{x}{T} \,,\label{newvars} \ee we find
\be I(T,0)= \frac{1}{4\pi^2} \frac{e^{-2imT}}{\sqrt{2m}\,T^{3/2}} \,\int^{\infty}_0 \sqrt{x}\, \frac{\sqrt{2+(x/2mT)}}{1+x/2mT}\,e^{-ix} \,dx \,.  \label{kernel2eqmass}\ee The leading contribution in the asymptotic long time limit is obtained by taking $T\rightarrow \infty$ inside the integral, from which we obtain the long time behavior
\be I(T,0) = -\frac{1}{4\pi^2}\,\sqrt{\frac{\pi}{2}}\,\frac{e^{-2imT}\,e^{i\frac{\pi}{4}}}{\sqrt{2m}\,T^{3/2}} + \cdots \label{asy2eqmass}\ee Using a similar change of variables, we find the general result that if the
spectral density vanishes at threshold as
\be \rho(K_0,K) \simeq \Big[K_0-K_T\Big]^{\alpha} ~~;~~ K_0 \rightarrow K_T \,,\label{rhovanish} \ee then the leading long time asymptotic behavior is
\be I(T,K) \propto \frac{1}{K^\alpha_T\, T^{\alpha+1}}\,.  \label{asygeneral}\ee

\begin{figure}[!h]
\begin{center}
\includegraphics[height=4in,width=4.5in,keepaspectratio=true]{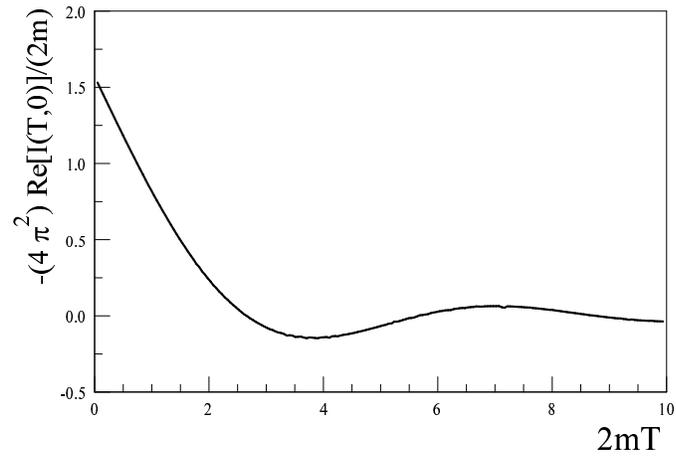}
\caption{  $-(4\pi^2)\,\mathrm{Re}I[T,0]/(2m)$ vs $2mT$ for $m_3=m_4,k=0$.}
\label{fig:realkernel}
\end{center}
\end{figure}

\begin{figure}[!h]
\begin{center}
\includegraphics[height=4in,width=4.5in,keepaspectratio=true]{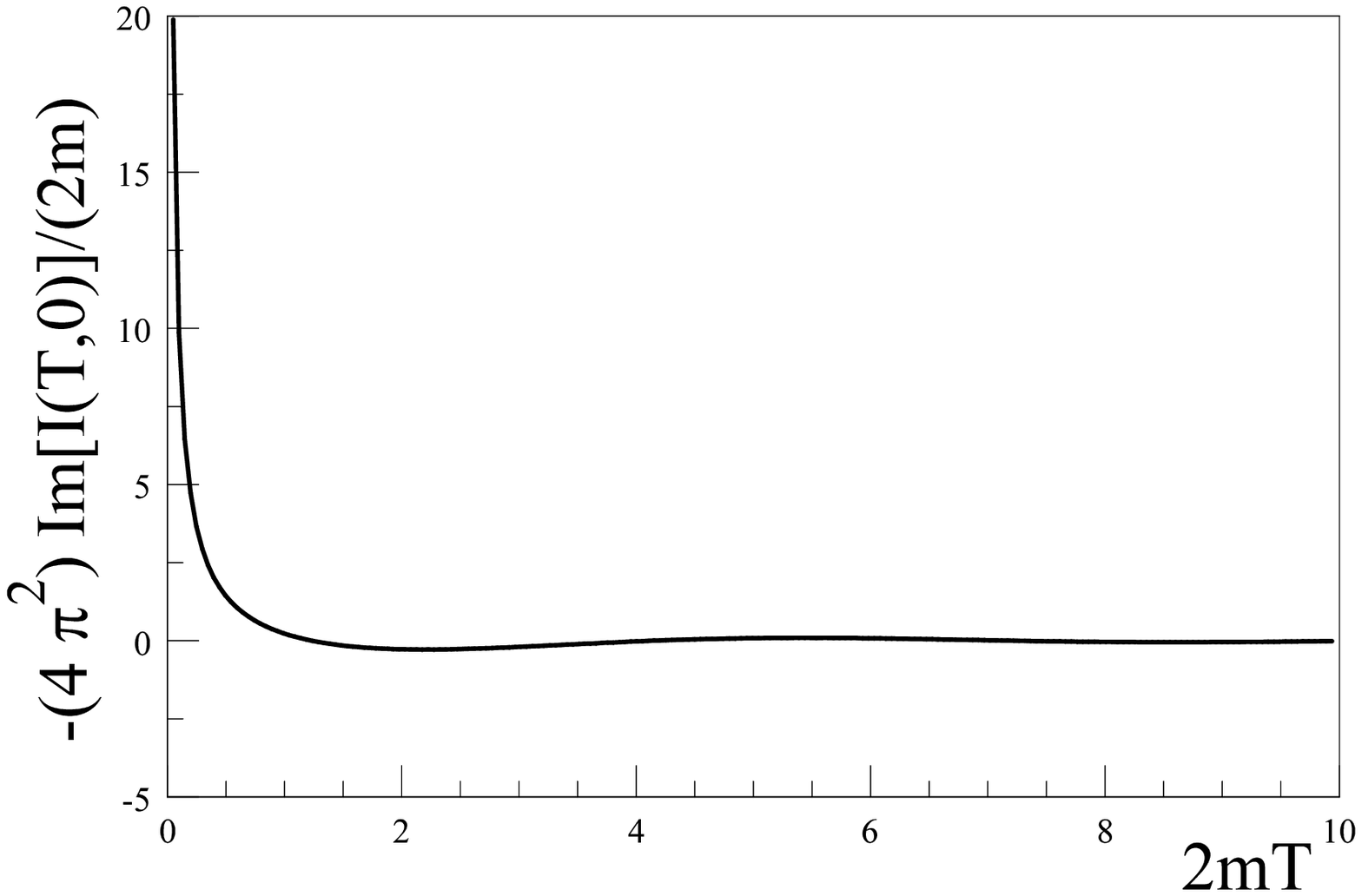}
\caption{  $-(4\pi^2) \,\mathrm{Im}I[T,0]/(2m)$ vs $2mT$ for $m_3=m_4,k=0$.}
\label{fig:imagkernel}
\end{center}
\end{figure}

\begin{figure}[!h]
\begin{center}
\includegraphics[height=4in,width=4.5in,keepaspectratio=true]{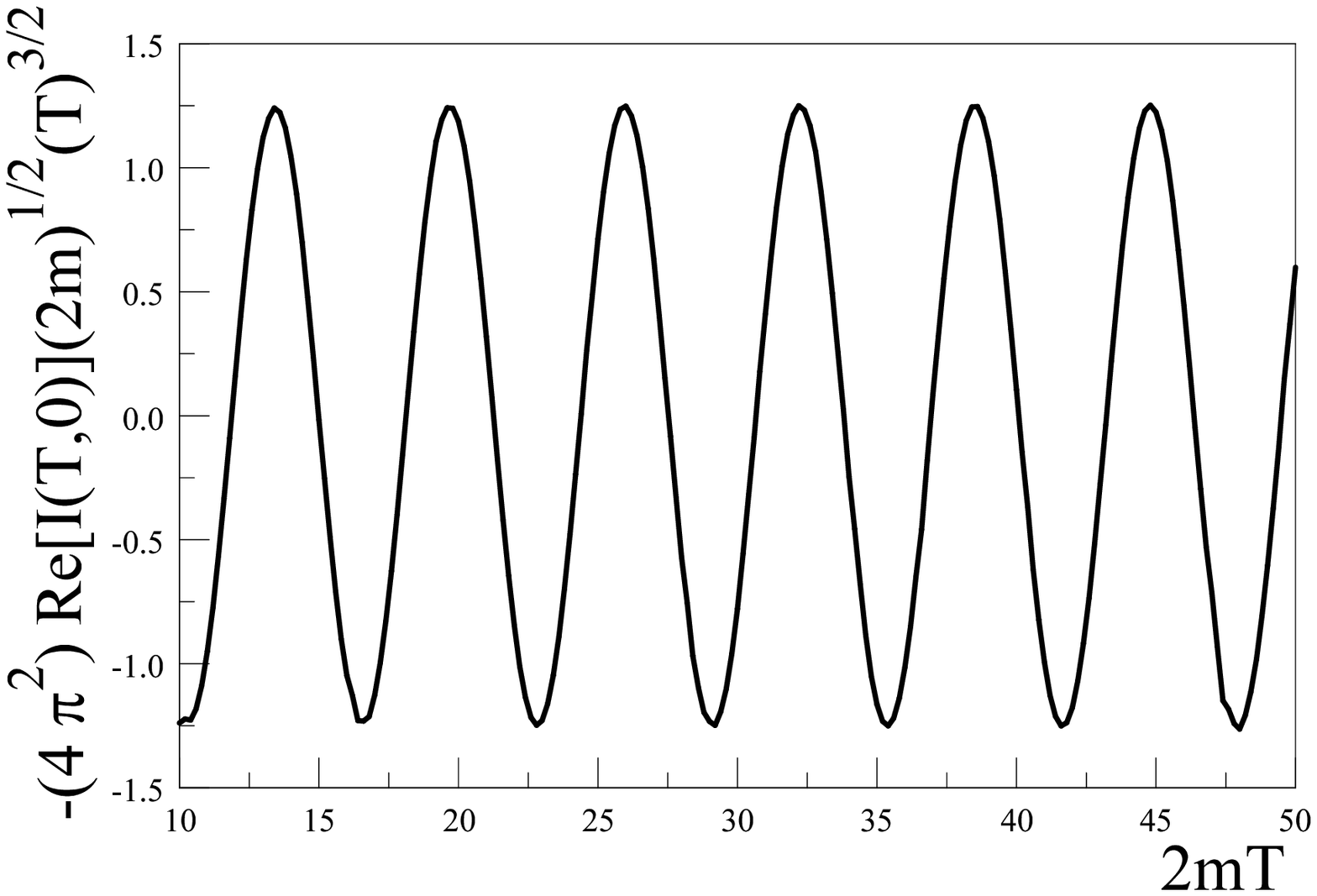}
\caption{  $-(4\pi^2)\,\mathrm{Re}I[T,0](2m)^{1/2}\,T^{3/2}$ vs $2mT$ for $m_3=m_4,k=0$.}
\label{fig:realkernel32}
\end{center}
\end{figure}

\begin{figure}[!h]
\begin{center}
\includegraphics[height=4in,width=4.5in,keepaspectratio=true]{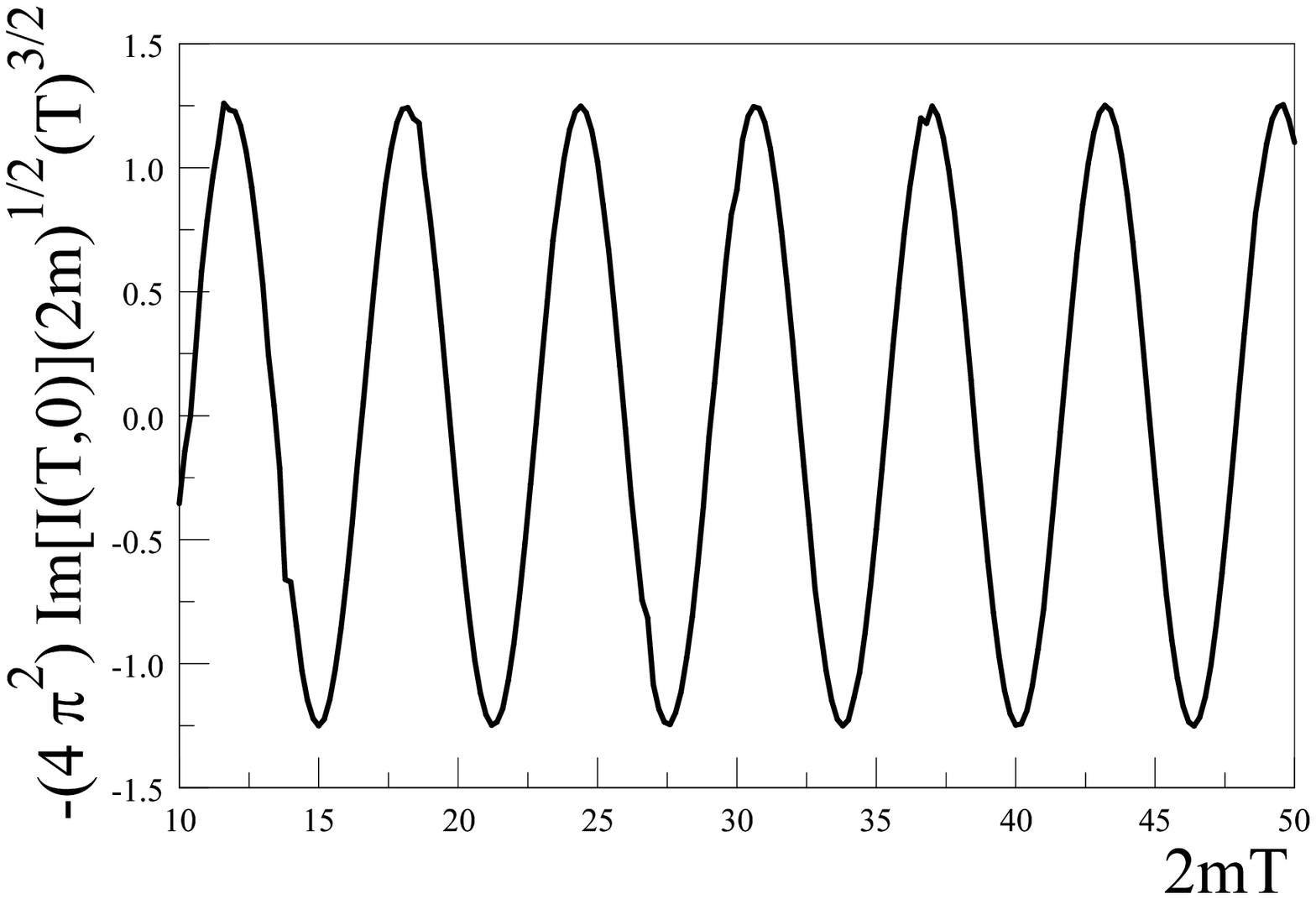}
\caption{  $-(4\pi^2)\,\mathrm{Im}I[T,0](2m)^{1/2}\,T^{3/2}$ vs $2mT$ for $m_3=m_4,K=0$.}
\label{fig:imagkernel32}
\end{center}
\end{figure}

Figs. (\ref{fig:realkernel},\ref{fig:imagkernel}) show  the $\mathrm{Re}I(T,0)$ and $\mathrm{Im}I(T,0)$ respectively for the   case of two equal masses in the final state, with spectral density (\ref{rho2masseq}), exhibiting the $1/T$ behavior at short time, whereas Figs. (\ref{fig:realkernel32},\ref{fig:imagkernel32}) show  the $\mathrm{Re}I(T,0) \,T^{3/2}$ and $\mathrm{Im}I(T,0)\,T^{3/2}$ in the long time regime confirming the scaling $1/T^{3/2}$. The result (\ref{asy2eqmass}) yields the value $\sqrt{\pi/2}=1.253$ for the amplitude of the oscillations in these figures, a result that is confirmed numerically. Furthermore, we find analytically that
\be -\frac{(4\pi)^2}{2m}\,\mathrm{Re}[I(T\rightarrow 0,0)] = \frac{\pi}{2} = 1.571 \label{zeroT} \ee which is confirmed numerically and displayed in fig. (\ref{fig:realkernel}).

This analysis can be generalized: let us rewrite the spectral density as
\begin{equation}
\rho(K_{0},K)=(K_{0}-K_{T})^{\alpha}\,\mathcal{F}\big[K_0;K\big]
\end{equation}

where $\mathcal{F}(K_{T};k)\neq0$.

Now in the spectral representation of the time kernel eqn. (\ref{specrep})
let us do the substitution, $K_{0}=K_{T}+\frac{x}{T}$, thus the integral
becomes
\begin{equation}
I=\frac{e^{-iK_{T}T}}{T^{1+\alpha}}\int \, x^{\alpha}\, e^{-ix} \,\mathcal{F}\big[K_{T}+\frac{x}{T},K\big]\,dx\,.
\end{equation}

The leading contribution in  the  $T\rightarrow\infty$ limit is given by

\begin{equation}
I=\frac{e^{-iK_{T}T}}{T^{1+\alpha}}\mathcal{F}\big[K_{T},k\big] \int_{0}^{\infty}dx\,x^{\alpha}\,e^{-ix}
\end{equation}

This integral can be regulated by introducing
\begin{equation}\int_{0}^{\infty}dx\,x^{\alpha}\, e^{-ix} \rightarrow \mathcal{J}(\alpha)\equiv \lim_{\epsilon\rightarrow0^{+}}\int_{0}^{\infty}dx\,x^{\alpha}\,e^{-ix(1-i\epsilon)} \,,
\end{equation}
with the values
\begin{equation}
\mathcal{J}(0)=-i;\quad \mathcal{J}(1/2)=- {e^{i\pi/4}}~\frac{\sqrt{\pi}}{2};\quad \mathcal{J}(1)=-1\,.
\end{equation}

\end{document}